# Latest progress on the reduced-order particle-in-cell scheme: II. Quasi-3D implementation and verification

M. Reza*[1], F. Faraji*, A. Knoll*

* Plasma Propulsion Laboratory, Department of Aeronautics, Imperial College London, London, United Kingdom

**Abstract**: Across many plasma applications, the underlying phenomena and interactions among the involved processes are known to exhibit three-dimensional characteristics. Furthermore, the global properties and evolution of plasma systems are often determined by a process called inverse energy cascade, where kinetic plasma processes at the microscopic scale interact and lead to macroscopic coherent structures. These structures can have a major impact on the stability of plasma discharges, with detrimental effects on the operation and performance of plasma technologies. Kinetic particle-in-cell (PIC) methods offer a sufficient level of fidelity to capture these processes and behaviors. However, three-dimensional PIC simulations that can cost-effectively overcome the curse of dimensionality and enable full-scale simulations of real-world time significance have remained elusive. Tackling the enormous computational cost issue associated with conventional PIC schemes, the computationally efficient reduced-order (RO) PIC approach provides a viable path to 3D simulations of real-size plasma systems. This part II paper builds upon the improvements to the RO-PIC's underpinning formulation discussed in part I and extends the novel "first-order" RO-PIC formulation to 3D. The resulting Quasi-3D (Q3D) implementation is rigorously verified in this paper, both at the module level of the Q3D reduced-dimension Poisson solver (RDPS) and at the global PIC code level. The plasma test cases employed correspond to 3D versions of the 2D configurations studied in Part I, specifically: **(a)** a 3D problem involving electron plasma oscillations with Landau damping, and **(b)** a 3D extension to the Diocotron instability problem. The detailed verifications of the Q3D RO-PIC confirm that it maintains the expected levels of cost-efficiency and accuracy, demonstrating the ability of the approach to indistinguishably reproduce full-3D simulation results at a fraction of the computational cost.

**Section 1: Introduction**

In many plasma applications of industrial relevance – ranging from magnetrons for material processing and Hall thrusters for spacecraft propulsion to magnetic confinement devices for fusion – the plasma dynamics is highly complex, primarily characterized by being three-dimensional in nature and exhibiting behaviors that extend across a vast range of temporal and spatial scales. A comprehensive understanding of the multi-dimensional and intricately coupled processes underlying plasma systems enables the prediction and control of plasmas, paving the way for more efficient technological solutions and novel applications, as well as harnessing fusion as the sustainable energy source for the future.

Nonetheless, our efforts toward acquiring an all-round knowledge of the three-dimensional plasma phenomena are in most cases impeded by the unavailability of a computationally viable self-consistent, high-fidelity model. Influential plasma phenomena in many technological plasmas have a kinetic nature, meaning that they stem from processes that alters the velocity distribution functions of the plasma species. Accordingly, high-fidelity kinetic simulations, such as the particle-in-cell (PIC) method [1][2], are the desired tools to study and predict the plasmas evolution. However, the substantial computational cost of traditional multi-dimensional PIC codes renders simulating real-size devices over timescales of real-world significance unfeasible, limiting our studies to simplified and/or reduced-size geometries for relatively short timeframes.

As a tangible example, a single 10 $\mu s$-long 3D simulation, performed by Villafana et al. [3] to study the dynamics of high-frequency instabilities in a geometry representative of a medium-sized Hall thruster, incurred a computational cost of $1.4 \times 10^6$ CPU hours. This enormous computational demand for such a short-duration 3D simulation underscores the unsuitability of traditional PIC implementations for practical purposes.

There is also an additional complexity. Previous research [4]-[13] has highlighted that plasma processes, such as the microscopic and macroscopic instabilities, as well as the plasma particles' dynamics and interactions with the imposed electromagnetic fields, are significantly influenced by the operating regime and physical parameters of the plasma system. Hence, there is also the need for understanding these parametric variabilities. Nonetheless, such extensive parametric studies using conventional PIC codes are most often impractical due to the same issue of cost inefficiency.

From a scientific standpoint, a computationally efficient 3D PIC model allows for the exploration of long-standing questions in plasma physics, enhancing our understanding of multi-dimensional plasma processes, their interactions, and parametric dependencies. From an engineering perspective, such predictive model not only can play a fundamental role in streamlining the design and development of more efficient and reliable plasma technologies but also can open doors to innovative technological solutions [14][15]. In this regard, development of new modelling approaches to reduce the computational burden of conventional kinetic simulations while maintaining the fidelity of their predictions can significantly boost plasma science and engineering.

---

[1] **Corresponding Author** (m.reza20@imperial.ac.uk)

To this end, we developed in 2023 the reduced-order (RO) PIC scheme [16]-[18] as a promising solution to address the computational cost challenge of traditional PIC schemes. Based upon the innovative concept of a reduced-dimension (RD) computational grid, the RO-PIC drastically reduces the computational complexity of the problems to simulate while preserving essential physics, serving as an effective alternative to conventional PIC methods. The results from testing RO-PIC across a diverse set of 2D plasma configurations [19]-[21] demonstrated the its capability to faithfully reproduce the behaviors observed in full-2D simulations, thereby affirming its utility for both physics studies and engineering applications. The rigorous verifications and the detailed characterizations of the convergence behavior of the RO-PIC scheme in various controlled benchmark cases laid a solid foundation for the extension of the RO-PIC to 3D and further exploration of its application in more complex 3D plasma configurations.

In part I of this article [22], we introduced recent advancements in the 2D RO-PIC model, focusing on the revision of the method's underpinning formulation from the original zeroth-order one to first-order. We evaluated the performance of the upgraded RO-PIC with first-order formulation against the original zeroth-order implementation. Through comparisons across multiple test cases at the module level and global code level, we demonstrated that the first-order RO-PIC yields substantial improvements in computational efficiency and prediction accuracy, notably outperforming the zeroth-order version [22].

Motivated by the successful adaptation of the 2D electrostatic RO-PIC based on the first-order formulation and its verifications in part I [22], we now extend the method to 3D in this part II and examine the resulting code's performance in several 3D test cases. We will discuss the details of this extension and present the verification outcomes for the electrostatic 3D RO-PIC (or quasi-3D [Q3D] PIC), highlighting its computational advantages and accuracy metric. The newly developed quasi-3D code serves as the Q3D branch of PICxelsior [22].

**Section 2: Extension of RO-PIC to 3D configuration**

The general methodology behind RO-PIC and its underlying formulation, which were discussed in part I [22], are readily extensible to 3D. For the three-dimensional implementation of RO-PIC, the reduced-dimension (RD) grid concept described in 2D in Refs. [16][22] can be adapted into a quasi-3D (Q3D) grid system, as illustrated in Figure 1. The RD grid system enables increasing the cell size along two dimensions beyond the Debye length.

In the following, we focus on the first-order Q3D PIC implementation. For completeness, the description of the zeroth-order Q3D PIC is provided in Appendix A. However, only the first-order Q3D PIC will be subject to tests and verifications in this paper.

The computational domain is decomposed into $M \times N \times P$ cubic regions, with fine computational grids placed on the edges of these regions along each direction. This arrangement results in a grid structure consisting of interconnected arrays of 1D grids along the $x$, $y$ and $z$ axes. The number of 1D grids along the $x$, $y$ and $z$ axes are $(N+1) \times (P+1)$, $(M+1) \times (P+1)$ and $(N+1) \times (M+1)$, respectively. Each of the 1D grids is characterized by fine-resolution cells along its respective dimension, ensuring that cell sizes remain below the Debye length and, hence, satisfying the stability requirement of the explicit PIC schemes [1][2].

The total number of cells and the particles count scale with $O(N_r^2 N_i)$ compared to $O(N_i^3)$, where $N_r$ is the total number of regions, and $N_i$ is the number of fine grid cells per dimension. Since the dimensionality-reduction technique underpinning RO-PIC enables approximating the 3D solution with $N_r \ll N_i$, it yields remarkable computational gain compared to a typical traditional explicit, momentum-conserving full-3D PIC simulation.

Similar to the Q2D implementation of RO-PIC, the Q3D extension requires adjustments to functions that interact with the grid to ensure compatibility with the RD grid definition. This involves modifying the "scatter" function, which transfers particle data to the grid, as well as the "gather" function, that retrieves grid-based data back to the particle positions. Additionally, the Poisson's equation and, hence, the Poisson solver used in an electrostatic PIC code must be adapted to align with the RD grid configuration. The modified functions are discussed in the following subsections.

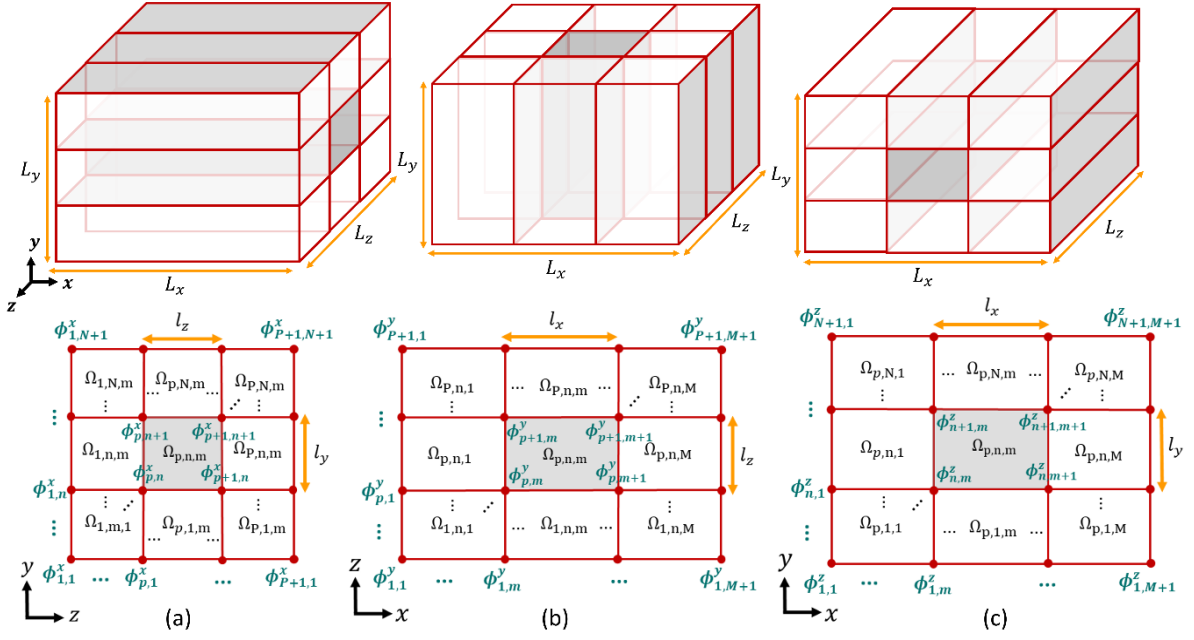

Figure 1: Schematic of Q3D grid and the domain decomposition into multiple regions for the first-order RO-PIC. The illustrations on the top row demonstrate the 3D views of domain decomposition along each coordinate plane. The illustrations on the bottom row show the cross-sectional views of the domain in different planes: (a) the $y - z$ plane, (b) the $z - x$ plane, and (c) the $y - x$ plane. The red dots represent the front views of the array of 1D fine grids along each corresponding direction placed along the edges of the cuboid subdomains.

## 2.1. Scatter and gather functions

In the scattering phase, the data associated with a particle, denoted by $\beta_s$, is deposited onto the Q3D computational grids. Note that subscript $s$ is used here to avoid indexing conflict with the index of regions along $z$ denoted by $p$. The deposition onto the grids is performed separately on each of the fine computational grids along the $x$, $y$ and $z$ directions

$$\beta_{p,n}^x(x_i) = \sum_s \beta_s W_{p,n,i}^x(\boldsymbol{r}_i, \boldsymbol{r}), \qquad \forall \boldsymbol{r}_i = (x_i, y_j, z_k) \in \Omega_{p,n,m} \tag{Eq. 1}$$

$$\beta_{p,m}^y(y_j) = \sum_s \beta_s W_{p,m,j}^y(\boldsymbol{r}_i, \boldsymbol{r}), \qquad \forall \boldsymbol{r}_i = (x_i, y_j, z_k) \in \Omega_{p,n,m} \tag{Eq. 2}$$

$$\beta_{n,m}^z(z_k) = \sum_s \beta_s W_{n,m,k}^z(\boldsymbol{r}_i, \boldsymbol{r}), \qquad \forall \boldsymbol{r}_i = (x_i, y_j, z_k) \in \Omega_{p,n,m} \tag{Eq. 3}$$

where, the summation is carried out over all simulation particles. $\boldsymbol{r}_i$ is the vector of grid node positions and $\boldsymbol{r}$ is a particle position vector. The 3D distribution of a property is then achieved by combining the values from the three grids as

$$\beta(x_i, y_j, z_k) = \frac{1}{3}\left(\beta_{p,n}^x(x_i) + \beta_{p,m}^y(y_j) + \beta_{n,m}^z(z_k)\right). \tag{Eq. 4}$$

In the gathering process, the grid data are interpolated at the particle's location. For each particle, the gathered value ($\beta_s(x, y, z)$) is a combination of the contributions from the three $x$-grid, $y$-grid, and $z$-grid, as per Eq. 5

$$\beta_s(x, y, z) = \frac{1}{3}\left(\sum_{p,n,i} \beta_{p,n}^x(x_i) W_{p,n,i}^x(\boldsymbol{r}_i, \boldsymbol{r}) + \sum_{p,m,j} \beta_{p,m}^y(y_j) W_{p,m,j}^y(\boldsymbol{r}_i, \boldsymbol{r}) \right.$$
$$\left. + \sum_{n,m,k} \beta_{n,m}^z(z_k) W_{n,m,k}^z(\boldsymbol{r}_i, \boldsymbol{r})\right); \quad \forall x, y, z \in \Omega_{p,n,m} \tag{Eq. 5}$$

Figure 2 provides a schematic to aid understanding the data scattering onto and its gathering from the Q3D grid. In Eqs. 1-5, $W_{p,n,i}^x$, $W_{p,m,j}^y$ and $W_{n,m,k}^z$ are the interpolation weight functions that determine how the data are interpolated between particles' positions and the neighboring grid nodes and vice versa. These weights depend on the relative position of the particle with respect to the grid nodes. The weighting functions also account for the core assumptions in the zeroth-order or first-order RO-PIC methods. In the first-order scheme, the weight functions prescribe a linear variation across the regions, allowing a smooth transition from one region to another

$$W_{p,n,i}^{x} = \begin{cases} \left(1 - \left|\frac{y - y_n}{l_y}\right|\right)\left(1 - \left|\frac{z - z_n}{l_z}\right|\right)\left(1 - \left|\frac{x - x_i}{\Delta x}\right|\right), & x, y, z \in \Omega_{p,n,m} \\ 0, & x, y, z \notin \Omega_{p,n,m} \end{cases}, \quad \text{(Eq. 6)}$$

$$W_{p,m,j}^{y} = \begin{cases} \left(1 - \left|\frac{z - z_n}{l_z}\right|\right)\left(1 - \left|\frac{x - x_m}{l_x}\right|\right)\left(1 - \left|\frac{y - y_j}{\Delta y}\right|\right), & x, y, z \in \Omega_{p,n,m} \\ 0, & x, y, z \notin \Omega_{p,n,m} \end{cases}, \quad \text{(Eq. 7)}$$

$$W_{n,m,k}^{z} = \begin{cases} \left(1 - \left|\frac{y - y_n}{l_y}\right|\right)\left(1 - \left|\frac{x - x_m}{l_x}\right|\right)\left(1 - \left|\frac{z - z_k}{\Delta z}\right|\right), & x, y, z \in \Omega_{p,n,m} \\ 0, & x, y, z \notin \Omega_{p,n,m} \end{cases}. \quad \text{(Eq. 8)}$$

As is discussed in Appendix A, in the zeroth-order scheme, each grid value is assumed to be constant perpendicular to the grid's direction across the regions.

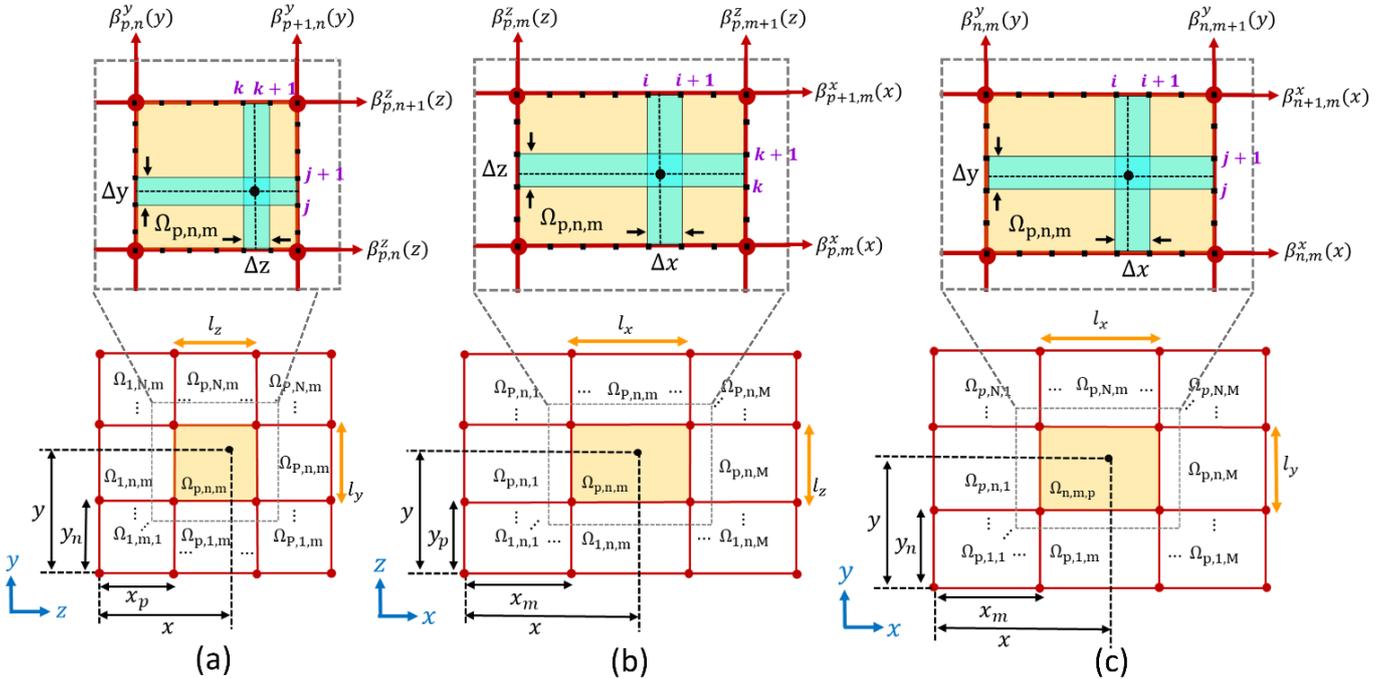

Figure 2: A schematic illustrating the exchange of data between the particles and reduced-dimension (RD) grid in (a) $y - z$ plane, (b) $z - x$ plane, and (c) $y - x$ plane.

### 2.2. Poisson solver

The reduced-dimension Poisson's/Gauss's equation introduced in part I [22] for the first-order implementation is extended to 3D here. This is achieved by writing the Gauss's law for the fine cells along the $x$-grid, $y$-grid and $z$-grid, separately. The computation cells along each direction and the electric flux through their surfaces are illustrated in Figure 3.

Considering $\boldsymbol{E} = -\boldsymbol{\nabla}\phi$, we arrive at the following equations, where the subscript $F$ stands for Front, $B$ for Back, $N$ for North, $S$ for South, $E$ for East, and $W$ for West of the computation cells

$$\frac{1}{\Delta x}\int_{z_B}^{z_F}\int_{y_S}^{y_N}\left(\frac{\partial \phi}{\partial x}\bigg|_{x_E} - \frac{\partial \phi}{\partial x}\bigg|_{x_W}\right)dydz + \int_{y_S}^{y_N}\left(\frac{\partial \phi}{\partial z}\bigg|_{z_F} - \frac{\partial \phi}{\partial z}\bigg|_{z_B}\right)dy + \int_{z_B}^{z_F}\left(\frac{\partial \phi}{\partial y}\bigg|_{y_N} - \frac{\partial \phi}{\partial y}\bigg|_{y_S}\right)dz$$
$$= \frac{1}{\epsilon_0}\int_{z_B}^{z_F}\int_{y_S}^{y_N}\rho\,dydz, \quad \text{(Eq. 9)}$$

$$\frac{1}{\Delta y}\int_{z_B}^{z_F}\int_{x_W}^{x_E}\left(\frac{\partial \phi}{\partial y}\bigg|_{y_N}-\frac{\partial \phi}{\partial y}\bigg|_{y_S}\right)dxdz+\int_{x_W}^{x_E}\left(\frac{\partial \phi}{\partial z}\bigg|_{z_F}-\frac{\partial \phi}{\partial z}\bigg|_{z_B}\right)dx+\int_{z_B}^{z_F}\left(\frac{\partial \phi}{\partial x}\bigg|_{x_E}-\frac{\partial \phi}{\partial x}\bigg|_{x_W}\right)dz$$
$$=\frac{1}{\epsilon_0}\int_{z_B}^{z_F}\int_{x_W}^{x_E}\rho\, dxdz,$$
(Eq. 10)

$$\frac{1}{\Delta z}\int_{y_S}^{y_N}\int_{x_W}^{x_E}\left(\frac{\partial \phi}{\partial z}\bigg|_{z_F}-\frac{\partial \phi}{\partial z}\bigg|_{z_B}\right)dxdy+\int_{x_W}^{x_E}\left(\frac{\partial \phi}{\partial y}\bigg|_{y_N}-\frac{\partial \phi}{\partial y}\bigg|_{y_S}\right)dx+\int_{y_S}^{y_N}\left(\frac{\partial \phi}{\partial x}\bigg|_{x_E}-\frac{\partial \phi}{\partial x}\bigg|_{x_W}\right)dy$$
$$=\frac{1}{\epsilon_0}\int_{y_S}^{y_N}\int_{x_W}^{x_E}\rho\, dxdy.$$
(Eq. 11)

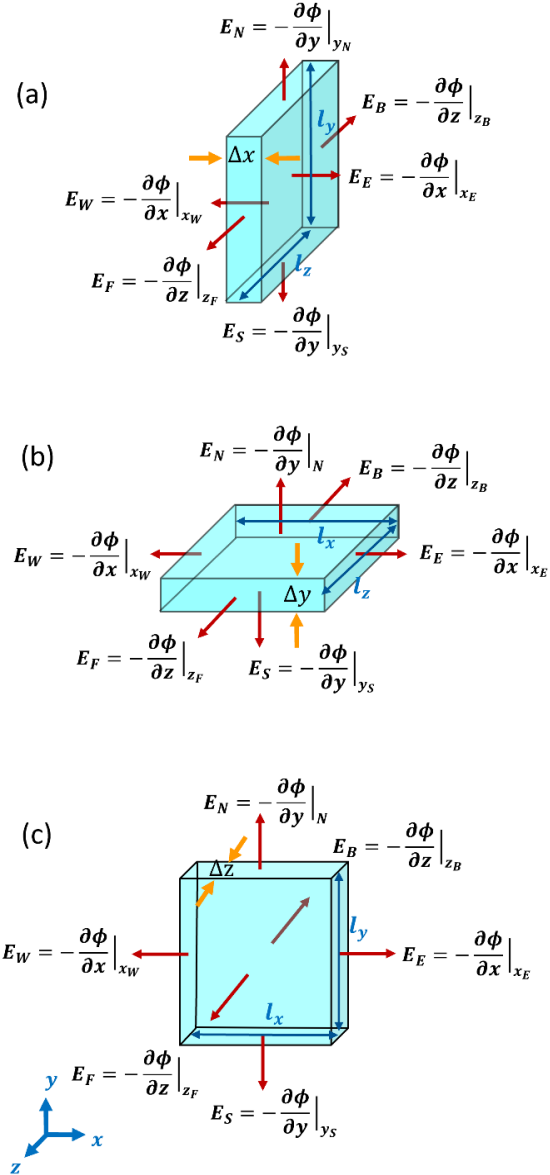

Figure 3: Schematic of the computational cells in region $\Omega_{p,n,m}$ associated with the dimensionality-reduction approach and the electric field components across their sides; (a) the cell along $x$-grid, (b) the cell along $y$-grid, and (c) the cell along $z$-grid.

At this point, we need to incorporate the core assumption of the RO-PIC method into Eqs. 9-11. Since the electric potential is solved on the reduced-dimension grid, we introduce the RO-PIC ansatz to specify how the potential at any given point $(x, y, z)$ in space is determined based on the partial potential values on the surrounding fine grids, i.e. $\phi^x(x)$, $\phi^y(y)$ and $\phi^z(z)$.

In the first-order formulation of RO-PIC, we assume that the potential varies linearly between adjacent 1D fine grids within a given region. Any arbitrary point in space is surrounded by four 1D grids along each spatial direction, resulting in a total of twelve 1D grids in 3D. Therefore, the potential at a given point, denoted by $\phi(x,y,z)$ is obtained through linear interpolation between these twelve surrounding 1D grids

$$\begin{aligned}\phi(x,y,z) \approx &\, (1-\tilde{y})(1-\tilde{z})\phi^x_{p,n}(x) + \tilde{y}(1-\tilde{z})\phi^x_{p,n+1}(x) + (1-\tilde{y})\tilde{z}\phi^x_{p+1,n}(x) + \tilde{y}\tilde{z}\phi^x_{p+1,n+1}(x) \\ &+ (1-\tilde{x})(1-\tilde{z})\phi^y_{p,m}(y) + \tilde{x}(1-\tilde{z})\phi^y_{p,m+1}(y) + (1-\tilde{x})\tilde{z}\phi^y_{p+1,m}(y) \\ &+ \tilde{x}\tilde{z}\phi^y_{p+1,m+1}(y) + (1-\tilde{x})(1-\tilde{y})\phi^z_{n,m}(z) + \tilde{x}(1-\tilde{y})\phi^z_{n,m+1}(z) \\ &+ (1-\tilde{x})\tilde{y}\phi^z_{n+1,m}(z) + \tilde{x}\tilde{y}\phi^z_{n+1,m+1}(z), \qquad \forall x,y,z \in \Omega_{p,n,m}.\end{aligned}$$

(Eq. 12)

where, $\tilde{x} = \frac{x-x_m}{l_x}$, $\tilde{y} = \frac{y-y_n}{l_y}$ and $\tilde{z} = \frac{z-z_p}{l_z}$.

Substituting Eq. 12 into Eqs. 9-11 yields the following coupled system of Poisson equations

$$\begin{aligned}&-\frac{l_y l_z}{64}\left[\frac{\partial^2 \phi^x_{p-1,n-1}}{\partial x^2} + \frac{\partial^2 \phi^x_{p-1,n+1}}{\partial x^2} + 6\frac{\partial^2 \phi^x_{p-1,n}}{\partial x^2} + 6\frac{\partial^2 \phi^x_{p,n-1}}{\partial x^2} + 36\frac{\partial^2 \phi^x_{p,n}}{\partial x^2} + 6\frac{\partial^2 \phi^x_{p,n+1}}{\partial x^2} + 6\frac{\partial^2 \phi^x_{p+1,n}}{\partial x^2} + \frac{\partial^2 \phi^x_{p+1,n-1}}{\partial x^2} + \frac{\partial^2 \phi^x_{p+1,n+1}}{\partial x^2}\right] \\ &- \frac{l_y}{8l_z}\left[\phi^x_{p-1,n-1} + \phi^x_{p-1,n+1} - 2\phi^x_{p,n-1} + 6\phi^x_{p-1,n} - 12\phi^x_{p,n} - 2\phi^x_{p,n+1} + 6\phi^x_{p+1,n} + \phi^x_{p+1,n-1} + \phi^x_{p+1,n+1}\right] \\ &- \frac{l_z}{8l_y}\left[\phi^x_{p-1,n-1} + \phi^x_{p+1,n-1} - 2\phi^x_{p-1,n} + 6\phi^x_{p,n-1} - 12\phi^x_{p,n} - 2\phi^x_{p+1,n} + 6\phi^x_{p,n+1} + \phi^x_{p-1,n+1} + \phi^x_{p+1,n+1}\right] \\ &- \frac{l_z}{8l_x}\int_{y_S}^{y_N}\left(\phi^y_{p-1,m} + \phi^y_{p-1,m+2} - 2\phi^y_{p-1,m+1} + 6\phi^y_{p,m} - 12\phi^y_{p,m+1} + 6\phi^y_{p,m+2} - 2\phi^y_{p+1,m+1} + \phi^y_{p+1,m} + \phi^y_{p+1,m+2}\right) dy \\ &- \frac{l_y}{8l_x}\int_{z_B}^{z_F}\left(\phi^z_{n-1,m} + \phi^z_{n-1,m+2} - 2\phi^z_{n-1,m+1} + 6\phi^z_{n,m} - 12\phi^z_{n,m+1} + 6\phi^z_{n,m+2} - 2\phi^z_{n+1,m+1} + \phi^z_{n+1,m} + \phi^z_{n+1,m+2}\right) dz \\ &- \frac{l_y}{8}\left[(1-\tilde{x})\left(\frac{\partial^2 \phi^z_{n-1,m}}{\partial z^2} + 6\frac{\partial^2 \phi^z_{n,m}}{\partial z^2} + \frac{\partial^2 \phi^z_{n+1,m}}{\partial z^2}\right) + \tilde{x}\left(\frac{\partial^2 \phi^z_{n-1,m+1}}{\partial z^2} + 6\frac{\partial^2 \phi^z_{n,m+1}}{\partial z^2} + \frac{\partial^2 \phi^z_{n+1,m+1}}{\partial z^2}\right)\right]_{z_B}^{z_F} \\ &- \frac{l_z}{8}\left[(1-\tilde{x})\left(\frac{\partial^2 \phi^y_{p-1,m}}{\partial y^2} + 6\frac{\partial^2 \phi^y_{p,m}}{\partial y^2} + \frac{\partial^2 \phi^y_{p+1,m}}{\partial y^2}\right) + \tilde{x}\left(\frac{\partial^2 \phi^y_{p-1,m+1}}{\partial y^2} + 6\frac{\partial^2 \phi^y_{p,m+1}}{\partial y^2} + \frac{\partial^2 \phi^y_{p+1,m+1}}{\partial y^2}\right)\right]_{y_S}^{y_N} \\ &- \frac{1-\tilde{x}}{l_z}\int_{y_S}^{y_N}\left(\phi^y_{p-1,m} - 2\phi^y_{p,m} + \phi^y_{p+1,m}\right) dy - \frac{\tilde{x}}{l_z}\int_{y_S}^{y_N}\left(\phi^y_{p-1,m+1} - 2\phi^y_{p,m+1} + \phi^y_{p+1,m+1}\right) dy \\ &- \frac{1-\tilde{x}}{l_y}\int_{z_B}^{z_F}\left(\phi^z_{n-1,m} - 2\phi^z_{n,m} + \phi^z_{n+1,m}\right) dz - \frac{\tilde{x}}{l_y}\int_{z_B}^{z_F}\left(\phi^z_{n-1,m+1} - 2\phi^z_{n,m+1} + \phi^z_{n+1,m+1}\right) dz = \frac{1}{\epsilon_0}\int_{z_B}^{z_F}\int_{y_S}^{y_N} \rho\, dy dz\end{aligned}$$

(Eq. 13)

$$\begin{aligned}&-\frac{l_x l_z}{64}\left[\frac{\partial^2 \phi^y_{p-1,m-1}}{\partial y^2} + \frac{\partial^2 \phi^y_{p-1,m+1}}{\partial y^2} + 6\frac{\partial^2 \phi^y_{p-1,m}}{\partial y^2} + 6\frac{\partial^2 \phi^y_{p,m-1}}{\partial y^2} + 36\frac{\partial^2 \phi^y_{p,m}}{\partial y^2} + 6\frac{\partial^2 \phi^y_{p,m+1}}{\partial y^2} + 6\frac{\partial^2 \phi^y_{p+1,m}}{\partial y^2} + \frac{\partial^2 \phi^y_{p+1,m-1}}{\partial y^2} + \frac{\partial^2 \phi^y_{p+1,m+1}}{\partial y^2}\right] \\ &- \frac{l_x}{8l_z}\left[\phi^y_{p-1,m-1} + \phi^y_{p-1,m+1} - 2\phi^y_{p,m-1} + 6\phi^y_{p-1,m} - 12\phi^y_{p,m} - 2\phi^y_{p,m+1} + 6\phi^y_{p+1,m} + \phi^y_{p+1,m-1} + \phi^y_{p+1,m+1}\right] \\ &- \frac{l_z}{8l_x}\left[\phi^y_{p-1,m-1} + \phi^y_{p+1,m-1} - 2\phi^y_{p-1,m} + 6\phi^y_{p,m-1} - 12\phi^y_{p,m} - 2\phi^y_{p+1,m} + 6\phi^y_{p,m+1} + \phi^y_{p-1,m+1} + \phi^y_{p+1,m+1}\right] \\ &- \frac{l_z}{8l_y}\int_{x_W}^{x_E}\left(\phi^x_{p-1,n} + \phi^x_{p-1,n+2} - 2\phi^x_{p-1,n+1} + 6\phi^x_{p,n} - 12\phi^x_{p,n+1} + 6\phi^x_{p,n+2} - 2\phi^x_{p+1,n+1} + \phi^x_{p+1,n} + \phi^x_{p+1,n+2}\right) dx \\ &- \frac{l_x}{8l_y}\int_{z_B}^{z_F}\left(\phi^z_{n,m-1} + \phi^z_{n+2,m-1} - 2\phi^z_{n+1,m-1} + 6\phi^z_{n,m} - 12\phi^z_{n+1,m} + 6\phi^z_{n+2,m} - 2\phi^z_{n+1,m+1} + \phi^z_{n,m+1} + \phi^z_{n+2,m+1}\right) dz \\ &- \frac{l_x}{8}\left[(1-\tilde{y})\left(\frac{\partial^2 \phi^z_{n,m-1}}{\partial z^2} + 6\frac{\partial^2 \phi^z_{n,m}}{\partial z^2} + \frac{\partial^2 \phi^z_{n,m+1}}{\partial z^2}\right) + \tilde{y}\left(\frac{\partial^2 \phi^z_{n+1,m-1}}{\partial z^2} + 6\frac{\partial^2 \phi^z_{n+1,m}}{\partial z^2} + \frac{\partial^2 \phi^z_{n+1,m+1}}{\partial z^2}\right)\right]_{z_B}^{z_F} \\ &- \frac{l_z}{8}\left[(1-\tilde{y})\left(\frac{\partial^2 \phi^x_{p-1,n}}{\partial x^2} + 6\frac{\partial^2 \phi^x_{p,n}}{\partial x^2} + \frac{\partial^2 \phi^x_{p+1,n}}{\partial x^2}\right) + \tilde{y}\left(\frac{\partial^2 \phi^x_{p-1,n+1}}{\partial x^2} + 6\frac{\partial^2 \phi^x_{p,n+1}}{\partial x^2} + \frac{\partial^2 \phi^x_{p+1,n+1}}{\partial x^2}\right)\right]_{x_W}^{x_E} \\ &- \frac{1-\tilde{y}}{l_z}\int_{x_W}^{x_E}\left(\phi^x_{p-1,n} - 2\phi^x_{p,n} + \phi^x_{p+1,n}\right) dx - \frac{\tilde{y}}{l_z}\int_{x_W}^{x_E}\left(\phi^x_{p-1,n+1} - 2\phi^x_{p,n+1} + \phi^x_{p+1,n+1}\right) dx \\ &- \frac{1-\tilde{y}}{l_x}\int_{z_B}^{z_F}\left(\phi^z_{n,m-1} - 2\phi^z_{n,m} + \phi^z_{n,m+1}\right) dz - \frac{\tilde{y}}{l_x}\int_{z_B}^{z_F}\left(\phi^z_{n+1,m-1} - 2\phi^z_{n+1,m} + \phi^z_{n+1,m+1}\right) dz = \frac{1}{\epsilon_0}\int_{z_B}^{z_F}\int_{x_W}^{x_E} \rho\, dx dz\end{aligned}$$

(Eq. 14)

$$\begin{aligned}
&-\frac{l_x l_y}{64}\left[\frac{\partial^2 \phi^z_{n-1,m-1}}{\partial z^2} + \frac{\partial^2 \phi^z_{n-1,m+1}}{\partial z^2} + 6\frac{\partial^2 \phi^z_{n-1,m}}{\partial z^2} + 6\frac{\partial^2 \phi^z_{n,m-1}}{\partial z^2} + 36\frac{\partial^2 \phi^z_{n,m}}{\partial z^2} + 6\frac{\partial^2 \phi^z_{n,m+1}}{\partial z^2} + 6\frac{\partial^2 \phi^z_{n+1,m}}{\partial z^2} + \frac{\partial^2 \phi^z_{n+1,m-1}}{\partial z^2} + \frac{\partial^2 \phi^z_{n+1,m+1}}{\partial z^2}\right] \\
&-\frac{l_x}{8l_y}\left[\phi^z_{n-1,m-1} + \phi^z_{n-1,m+1} - 2\phi^z_{n,m-1} + 6\phi^z_{n-1,m} - 12\phi^z_{n,m} - 2\phi^z_{n,m+1} + 6\phi^z_{n+1,m} + \phi^z_{n+1,m-1} + \phi^z_{n+1,m+1}\right] \\
&-\frac{l_y}{8l_x}\left[\phi^z_{n-1,m-1} + \phi^z_{n+1,m-1} - 2\phi^z_{n-1,m} + 6\phi^z_{n,m-1} - 12\phi^z_{n,m} - 2\phi^z_{n+1,m} + 6\phi^z_{n,m+1} + \phi^z_{n-1,m+1} + \phi^z_{n+1,m+1}\right] \\
&-\frac{l_y}{8l_z}\int_{x_W}^{x_E}(\phi^x_{p,n-1} + \phi^x_{p+2,n-1} - 2\phi^x_{p+1,n-1} + 6\phi^x_{p,n} - 12\phi^x_{p+1,n} + 6\phi^x_{p+2,n} - 2\phi^x_{p+1,n+1} + \phi^x_{p,n+1} + \phi^x_{p+2,n+1})dx \\
&-\frac{l_x}{8l_z}\int_{y_S}^{y_N}(\phi^y_{p,m-1} + \phi^y_{p+2,m-1} - 2\phi^y_{p+1,m-1} + 6\phi^y_{p,m} - 12\phi^y_{p+1,m} + 6\phi^y_{p+2,m} - 2\phi^y_{p+1,m+1} + \phi^y_{p,m+1} + \phi^y_{p+2,m+1})dy \\
&-\frac{l_x}{8}\left[(1-\tilde{z})\left(\frac{\partial^2 \phi^y_{p,m-1}}{\partial y^2} + 6\frac{\partial^2 \phi^y_{p,m}}{\partial y^2} + \frac{\partial^2 \phi^y_{p,m+1}}{\partial y^2}\right) + \tilde{z}\left(\frac{\partial^2 \phi^y_{p+1,m-1}}{\partial y^2} + 6\frac{\partial^2 \phi^y_{p+1,m}}{\partial y^2} + \frac{\partial^2 \phi^y_{p+1,m+1}}{\partial y^2}\right)\right]_{y_S}^{y_N} \\
&-\frac{l_y}{8}\left[(1-\tilde{z})\left(\frac{\partial^2 \phi^x_{p,n-1}}{\partial x^2} + 6\frac{\partial^2 \phi^x_{p,n}}{\partial x^2} + \frac{\partial^2 \phi^x_{p,n+1}}{\partial x^2}\right) + \tilde{z}\left(\frac{\partial^2 \phi^x_{p+1,n-1}}{\partial x^2} + 6\frac{\partial^2 \phi^x_{p+1,n}}{\partial x^2} + \frac{\partial^2 \phi^x_{p+1,n+1}}{\partial x^2}\right)\right]_{x_W}^{x_E} \\
&-\frac{1-\tilde{z}}{l_y}\int_{x_W}^{x_E}(\phi^x_{p,n-1} - 2\phi^x_{p,n} + \phi^x_{p,n+1})\,dx - \frac{\tilde{z}}{l_y}\int_{x_W}^{x_E}(\phi^x_{p+1,n-1} - 2\phi^x_{p+1,n} + \phi^x_{p+1,n+1})\,dx \\
&-\frac{1-\tilde{z}}{l_x}\int_{y_S}^{y_N}(\phi^y_{p,m-1} - 2\phi^y_{p,m} + \phi^y_{p,m+1})\,dy - \frac{\tilde{z}}{l_x}\int_{y_S}^{y_N}(\phi^y_{p+1,m-1} - 2\phi^y_{p+1,m} + \phi^y_{p+1,m+1})\,dy = \frac{1}{\epsilon_0}\int_{y_S}^{y_N}\int_{x_W}^{x_E}\rho\,dxdy
\end{aligned}$$
(Eq. 15)

Eqs. 13-15 represent the first-order Q3D approximation of the Poisson's equation, which are numerically solved using the Q3D reduced-dimension Poisson solver (RDPS). RDPS employs Julia's efficient built-in direct matrix-solving algorithm based on the LU decomposition.

The zeroth-order formulation for the Q3D Poisson's system of equations is provided in Appendix A.

### Section 3: Standalone verifications of the first-order 3D RDPS

We present here the verification results of the 3D RDPS in two standalone Poisson problems, as defined in Table 1. In case 1, the boundary condition plays a dominant role in determining the solution, while in case 2, the charge distribution $\rho(x, y, z)$ has the main influence on the outcome.

Both test cases represent a Cartesian $(x - y - z)$ domain of unit length along each direction. 50 computational cells are used along $x$, $y$ and $z$ dimensions, thus $N_i = N_j = N_k = 50$. The permittivity of free space is assumed to be unity ($\epsilon_0 = 1$).

The RDPS' solutions with various number-of-regions for case 1 and case 2 are illustrated in Figure 4 and Figure 5, respectively. As the salient observation, similar to the solutions obtained from the first-order 2D RDPS in part I [22], the 3D RDPS' solutions in both test cases are smooth, and beyond 10 regions, they become visually indistinguishable from the full-3D results.

| Case No. | Source terms ($\rho(x, y, z)$) | Boundary Conditions |
|---|---|---|
| 1 | $\rho = 0$ | $\phi(0, y, z) = \phi(L_x, y, z) = \sin(\pi y)\sin(\pi z)$<br>$\phi(x, 0, z) = \phi(x, L_y, z) = 0$<br>$\phi(x, 0, z) = \phi(x, L_y, z) = \sin(3\pi x)\sin(\pi y)$ |
| 2 | $\rho = \sin(\pi r)\cos(3\Phi)\cos(2\theta_1)\cos(2\theta_2)\exp(-5r^2)$;<br>$r = \sqrt{(x-0.5)^2 + (y-0.5)^2 + (z-0.5)^2}$,<br>$\Phi = \tan^{-1}\left(\frac{\sqrt{(x-0.5)^2 + (y-0.5)^2}}{z - 0.5}\right)$,<br>$\theta_1 = \tan^{-1}\left(\frac{y - 0.5}{x - 0.5}\right)$,<br>$\theta_2 = \tan^{-1}\left(\frac{z - 0.5}{y - 0.5}\right)$. | $\phi(0, y, z) = \phi(L_x, y, z) = 0$<br>$\phi(x, 0, z) = \phi(x, L_y, z) = 0$<br>$\phi(x, y, 0) = \phi(x, y, L_y) = 0$ |

Table 1: Definition of the 3D Poisson problems for standalone verifications of the first-order reduced-dimension Poisson solver.

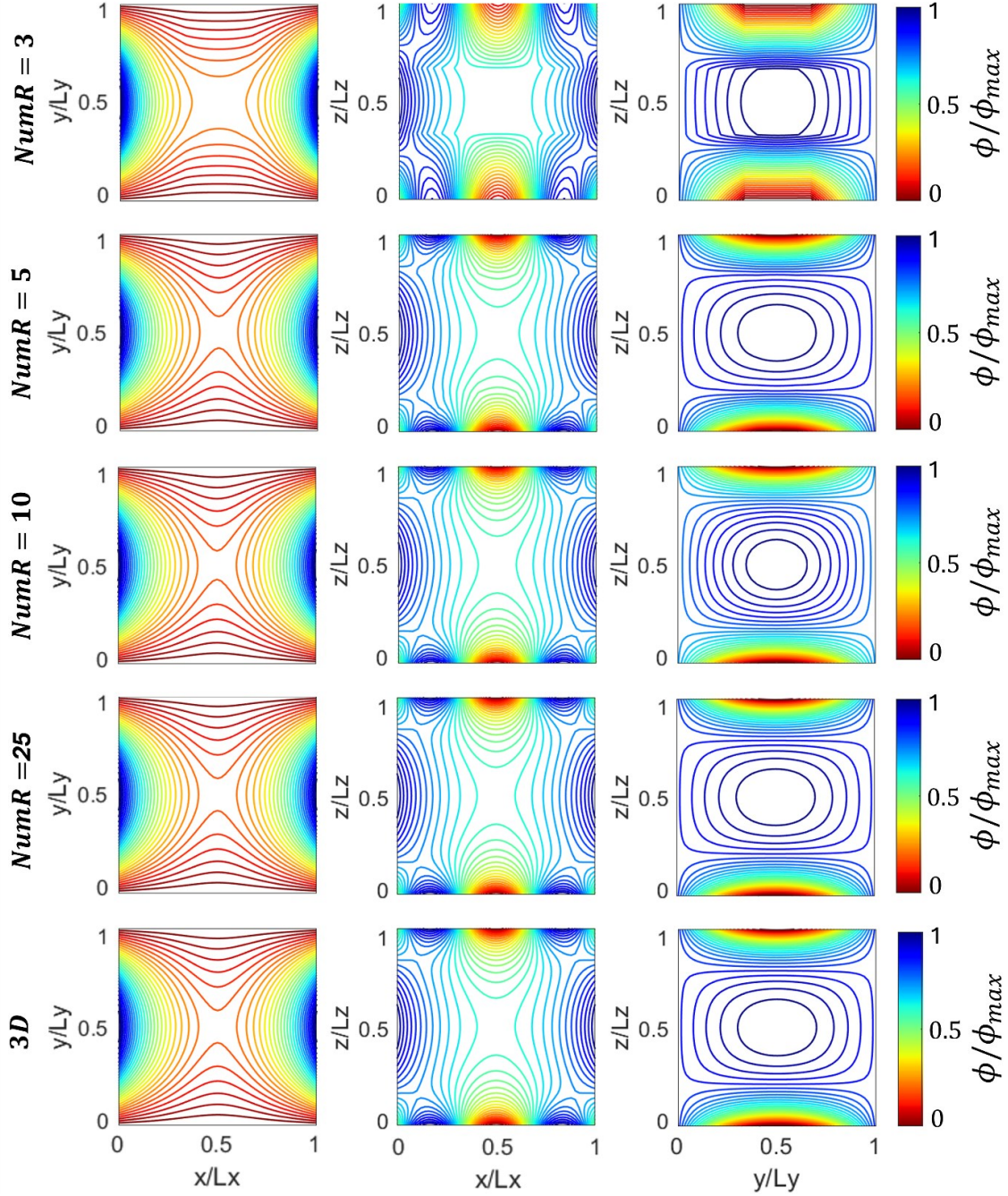

Figure 4: Comparison of the first-order RDPS' solutions with various number of regions against the corresponding full-3D solution (**bottom-most row**) for case 1 in the $y - x$ plane (**left column**), the $z - x$ plane (**middle column**), and $z - y$ plane (**right column**). The numbers of regions are equal along all three dimensions ($M = N = P$), which are denoted by *NumR*.

Figure 6 presents the error convergence plots of the reduced-order potential approximations $\phi_{Q3D}(i,j,k)$ for the defined Poisson problems across various number of regions, along with the corresponding ratio of the computational cells in Q3D to those in the full-3D Poisson solver. The error is computed according to Eq. 16 and with respect to the corresponding full-3D solution

$$Error = 100 \frac{\sqrt{\sum_{k=1}^{N_k} \sum_{i=1}^{N_i} \sum_{j=1}^{N_j} \left(\phi_{3D}(i,j,k) - \phi_{Q3D}(i,j,k)\right)^2}}{\sqrt{\sum_{k=1}^{N_k} \sum_{i=1}^{N_i} \sum_{j=1}^{N_j} \left(\phi_{3D}(i,j,k)\right)^2}}. \qquad \text{(Eq. 16)}$$

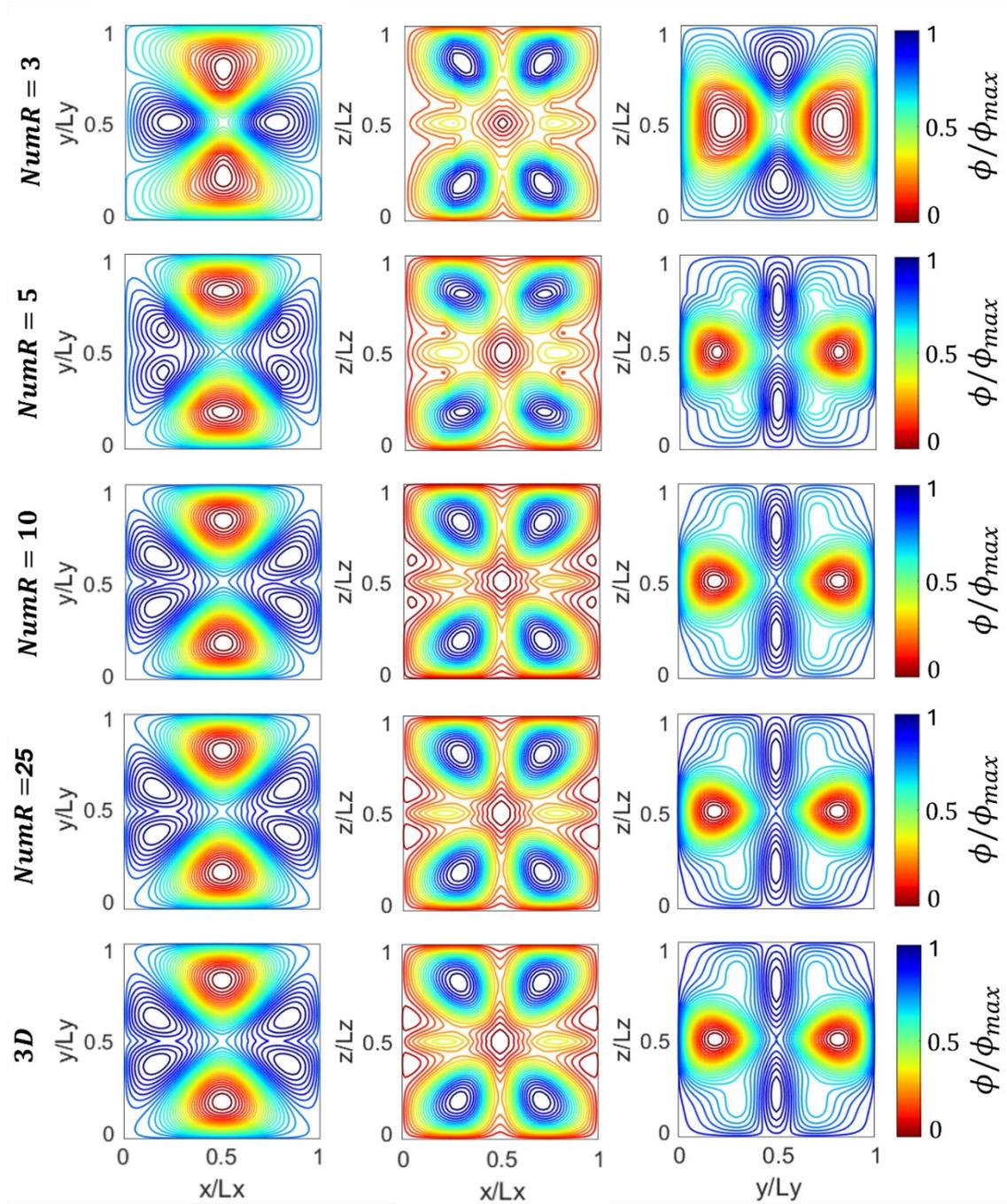

Figure 5: Comparison of the RDPS' solutions with various number of regions against the corresponding full-3D solution (**bottom-most row**) for case 2 in the $y - x$ plane (**left column**), the $z - x$ plane (**middle column**) and the $z - y$ plane (**right column**). The numbers of regions are equal along all three dimensions ($M = N = P$), which are denoted by *NumR*.

In both test cases, we observe from Figure 6 that the error of the Q3D potential approximation remains below 5 % with just 5 regions along each direction, resulting in the number of computational cells being only 3 % of the full 3D model.

Furthermore, with 10 regions, the errors in either test case drop to below 0.4 %, while the computational cells account for 10 % of those in the full-3D setup. It is worth noting that the ratio of Q3D to full-3D cells ($N_{cell,Q3D}/N_{cell,3D}$) is indicative of the computational gain resulted from using the reduced-dimension solver compared to the standard one.

The test cases presented here involved full Dirichlet boundary conditions. In Appendix B, results of the 3D RDPS verifications are shown for test cases that feature a variety of boundary conditions, including Neumann and periodic boundary conditions.

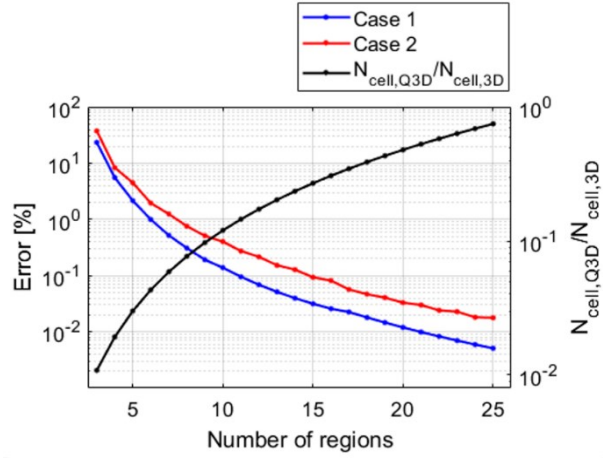

Figure 6: Error convergence of the RDPS' solution; variation of the normalized L2 error (Eq. 16) between the full-3D solution and the approximate Q3D solution from the RDPS when increasing the number of regions for test cases 1 and 2. The right axis represents the ratios of computational cells in Q3D to those in the full-3D for the respective number-of-region. The y-axes are both in logarithmic scale.

## Section 4: RO-PIC verifications

Following the verifications of the 3D RDPS in standalone Poisson test problems and ensuring the correctness of the underpinning 3D dimensionality-reduction formulation and its numerical implementation, the next step involves integration of the RDPS within the RO-PIC loop and verifying the complete Q3D code in plasma test cases.

The adopted test cases, which are discussed respectively in the following subsections, include: **(1)** an electrostatic setup involving electron plasma oscillations with Landau damping [23], and **(2)** 3D extension to the Diocotron instability which features an applied magnetic field [24][25]. These test cases are used to assess and characterize the convergence behavior of the 3D RO-PIC with varying number-of-regions.

### 4.1. Test case 1: Electron plasma oscillations

#### 4.1.1. Description of the problem setup

This test case is adapted from the example problem provided in Ref. [23] and is a 3D extension to the 2D plasma test case number 1 presented in Part I [22]. The 3D Cartesian $(x - y - z)$ computational domain of the problem is shown in Figure 7.

To initiate the simulations, the ions are uniformly distributed across the domain with a density of $n_0 = 1 \times 10^{11} m^{-3}$ and a zero initial temperature $T_{i,0} = 0$. Electrons, sampled from a Maxwellian distribution with a temperature of 0.1 eV and the density of $n_0 = 1 \times 10^{11} m^{-3}$, are loaded into an octant of the domain with $x \in [0, L_x/2)$, $y \in [0, L_y/2)$, $z \in [0, L_z/2)$.

Zero-volt Dirichlet boundary condition are applied on all boundaries for the potential and reflective boundary condition is imposed on particles crossing any of the domain boundaries. The main simulation parameters are detailed in Table 2.

The initial charge imbalance arises from the concentration of the electrons in one octant, while ions occupy the entire domain, generating an electric field. The electric field acts on the electrons to propagate outward to achieve charge neutrality, which in turn initiates the plasma oscillations. As the system evolves, these oscillations are damped by Landau damping [26], leading to a gradual restoration of quasi-neutrality as the electrons redistribute, ultimately establishing a more stable plasma state.

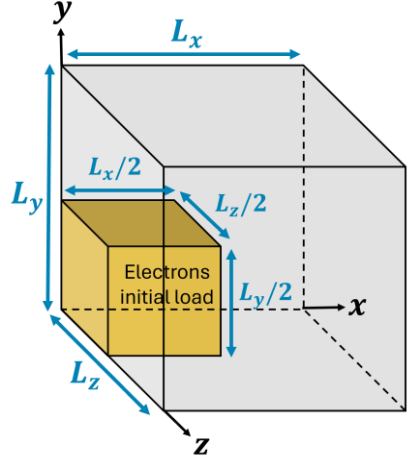

Figure 7: Schematic of the computational domain and setup for test case 1.

| Parameter | Value [unit] |
|---|---|
| Physical Parameters | |
| Initial plasma density ($n_0$) | $1 \times 10^{11}$ [m$^{-3}$] |
| Initial electron temperature ($T_{e,0}$) | 0.1 [eV] |
| Initial ion temperature ($T_{i,0}$) | 0.0 [eV] |
| Potential at walls ($\phi_w$) | 0 [V] |
| Ion mass | $2.65686 \times 10^{-26}$ [kg] |
| Computational Parameters | |
| Time step ($\Delta t$) | $1 \times 10^{-9}$ [s] |
| Time averaging for outputs | $1 \times 10^{-8}$ [s] |
| Total simulation time ($t_{sim}$) | 30 [μs] |
| Domain length ($L_x = L_y = L_z$) | 2 [cm] |
| Cell size ($\Delta x = \Delta y = \Delta z$) | $5 \times 10^{-2}$ [cm] |
| Number of computational nodes along each direction ($N_i = N_j = N_k$) | 40 |
| Initial number of macroparticles per cells ($N_{ppc}$) | 50 |

Table 2: Summary of the main computational and physical parameters used for first-order Q3D simulations of test case 1.

*4.1.2. Results*

Before delving into the discussion of the results from the Q3D simulations, it is important to highlight the impact of including the third dimension in this test case. This is provided in Appendix C.a, where a comparison is shown between the time evolutions of the electric potential energy ($E_{pot}$) and the particles' kinetic energy ($E_{kin}$) within the system from full-2D and full-3D simulations. The energies are computed according to Eqs. 2 and 3 in part I [22]. According to the figure in Appendix C, the 2D simulations show more pronounced oscillations, whereas the amplitude of the oscillations in 3D is smaller and is also damped more quickly.

Now, turning our focus to the Q3D results for test case 1, the evolution of energies from the reduced-order simulations with different number of regions and from the full-3D simulation are compared in Figure 8. In addition, Figure 9 illustrates sample plasma potential snapshots in different time instants during the plasma oscillations. The corresponding electron number density snapshots are presented in Appendix C.a.

Similar to the observations made in part I in two-dimensions [22], the overall behavior of the system in terms of the development of the plasma oscillations and the exchange between the electric potential and kinetic energy is properly captured in all Q3D simulations, though with varying accuracies depending on the number of regions used.

The variation in the accuracy of the Q3D simulations with the number of regions can also be visually assessed through the resolution and the level of detail captured in the snapshots of the electric potential and the electron density (Figure 9 and Appendix C.a, respectively).

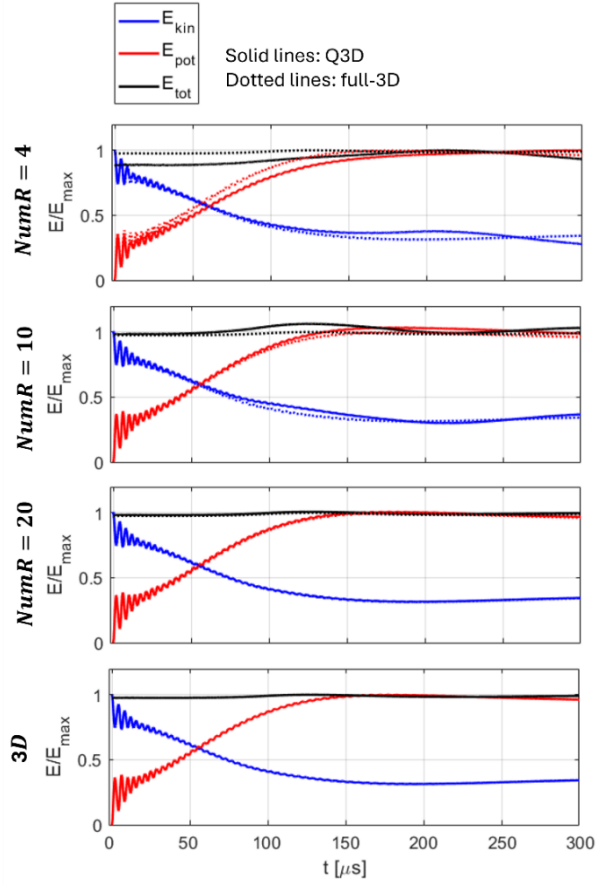

Figure 8: Time evolution of the normalized total electric potential energy ($E_{pot}$), the total electrons' kinetic energy ($E_{kin}$), and the total energy of the system ($E_{tot} = E_{pot} + E_{kin}$) from the Q3D and the full-3D simulations of test case 1. The results from the full-3D simulations are superimposed as dotted lines on the Q3D plots. The normalization factor ($E_{max}$) is the maximum of the respective energies from the full-3D simulation over the simulated duration.

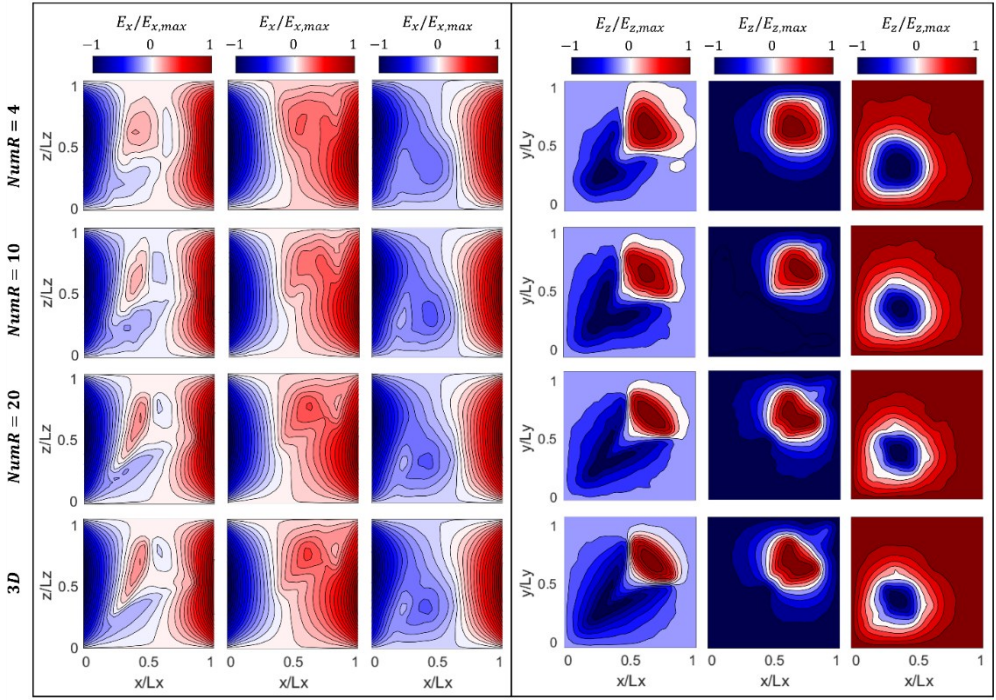

Figure 9: 2D snapshots from the Q3D and the full-3D simulations of test case 1 at three different time instants during the electron plasma oscillations; **left panel**: normalized electric field along $x$ ($E_x/E_{x,max}$) in the $z - x$ plane, **right panel**: normalized electric field along $z$ ($E_z/E_{z,max}$) in the $y - x$ plane. $E_{x,max}$ and $E_{z,max}$ are the peaks of the $E_x$ and the $E_z$ in the respective snapshot from the full-3D simulation.

The errors in the predicted plasma oscillations characteristics from the Q3D simulations, including the oscillations' frequency and damping rate, in addition to the associated computational cost reductions, are provided in Figure 10.

It is noticed that the error values and the decreasing error trend as the number of regions in the Q3D simulations increase are consistent with those observed in first-order 2D simulations, indicating that extending the approach to 3D does not affect the convergence behavior or accuracy level of the reduced-order approximation in this test case. Notably, with 10 regions, the oscillations characteristics are captured with an error margin of less than 1 %, while the computational time is reduced to only 5 % of that required for a full-3D simulation. This highlights the efficiency of the Q3D RO-PIC.

We would highlight that the presented analysis serves as a single test case and drawing a more robust and general conclusion about the similarity of RO-PIC's convergence behavior in 2D and 3D requires testing the approach across a wider range of cases. This is why we have introduced the second test case in the following subsection.

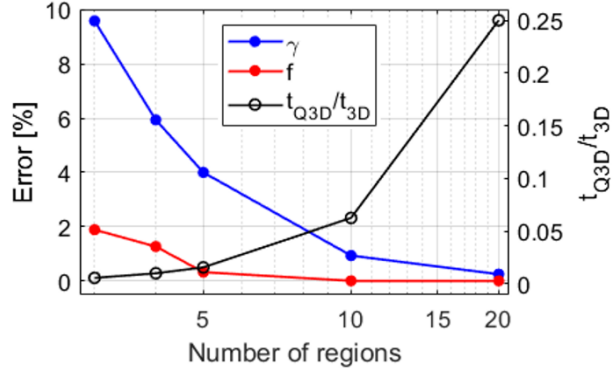

Figure 10: Variation of the error in the predictions of the frequency (f) and damping rate ($\gamma$) of oscillations in the electric potential energy ($E_{pot}$) from the Q3D simulations with various number of regions compared against the full-3D results in test case 1. The computational time ratio between the Q3D and the full-3D simulations is also plotted against the number of regions used in the Q3D simulations.

### 4.2. Test case 2: Diocotron instability

#### 4.2.1. Description of the problem setup

The Diocotron instability test case described in part I [22] is extended to 3D configuration here. The simulation thus simulates a 3D Cartesian domain, illustrated in Figure 11, with the dimensions $L_x = L_y = L_z = 10\lambda_D$, where $\lambda_D$ represents the Debye length.

To introduce 3D effects in the setup, upon the initialization of the simulation, the electrons are loaded only on the mid-plane of the domain (on the $x - y$ plane at $z = L_z/2$) following a ring-shaped Gaussian distribution with a radius $R_{load} = L_x/4$ centered at $x = L_x/2$ and $y = L_y/2$ and a standard deviation of $\sigma = 0.03L_x$. A slight perturbation is introduced to the azimuthal positions of the electrons on the Gaussian ring to accelerate the onset of instability. The ions are modeled as a uniform, stationary background with a constant density of $n_i = 1 \times 10^{12} \, m^{-3}$ over the load plane. The applied magnetic field is along $z$-direction with a uniform intensity of $B_z = 2.5 \, mT$. The numerical parameters used in the simulations are listed in Table 3. All domain boundaries are grounded with zero potential, and particles reaching the boundaries are specularly reflected into the domain.

As a brief overview of the physics underlying this test problem, in the presence of a magnetic field, the establishment of a radial electric field induces differential rotation within the plasma, leading to velocity gradients among different plasma elements. This differential rotation results in the amplification of small perturbations in the plasma and, hence, the excitation of the Diocotron instability [24][25]. This instability is characterized by the growth of azimuthal perturbations, resulting in the formation of vortex structures.

In the present configuration, the vortex structures predominantly emerge in the mid-plane of the domain, where the electrons are initially introduced, and the electric field is the strongest. As the instability evolves, the vortices grow and interact with each other, leading to complex dynamics within the plasma. In the meantime, as the perturbations develop, electrons disperse along the $z$-axis, which introduces a three-dimensional effect in the plasma's behavior. This axial dispersion creates variations in the electron density throughout the plasma volume, contributing to the complexity of the vortex structures.

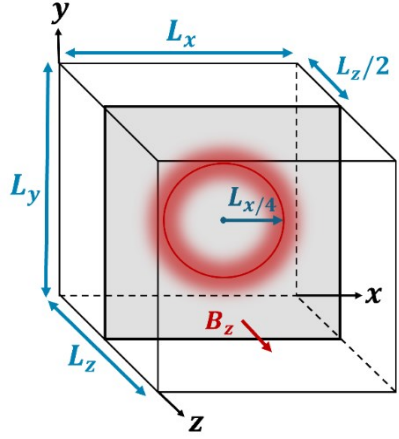

Figure 11: Schematic of the computational domain and setup of test case 2.

| Parameter | Value [unit] |
|---|---|
| Physical Parameters | |
| Initially loaded peak electron density ($n_e$) | $1 \times 10^{12}$ [m$^{-3}$] |
| Initial electron temperature ($T_{e,0}$) | 1 [eV] |
| Background ion density | $1 \times 10^{12}$ [m$^{-3}$] |
| Applied magnetic field ($B_z$) | 2.5 [mT] |
| Potential at walls ($\phi_w$) | 0 [V] |
| Computational Parameters | |
| Time step ($\Delta t$) | $0.1\omega_{pe}$ |
| Time averaging for outputs | $\omega_{pe}$ |
| Total simulation time ($t_{sim}$) | $100\omega_{pe}$ |
| Domain length ($L_x = L_y = L_z$) | $10\lambda_D$ |
| Cell size ($\Delta x = \Delta y = \Delta z$) | $0.1\lambda_D$ |
| Number of computational nodes along each direction ($N_i = N_j = N_k$) | 100 |
| Initial number of macroparticles per cells ($N_{ppc}$) | 100 |
| Initial maximum plasma frequency ($\omega_{pe}$) | $5.64 \times 10^7$ [rad/s] |
| Initial minimum Debye length ($\lambda_D$) | $7.43 \times 10^{-1}$ [cm] |

Table 3: Summary of the main computational and physical parameters used for simulations of test case 2.

### 4.2.2. Results

Prior to presenting the results from the Q3D simulations, the figure in Appendix C.b displays 3D snapshots of the plasma evolution at different time instants from a full-3D simulation. These snapshots illustrate the expected plasma behavior in this configuration, including simultaneous development of the Diocotron instability and the dispersion of the electrons across the z-axis.

Additionally, we conducted a 2D simulation under the conditions similar to those in the 3D simulation presented in the preceding section, with adjustments made to exclude the z-dimension. A comparative analysis of the time-averaged profiles of the plasma properties obtained from the 2D and 3D simulations is presented in the second figure of Appendix C.b, highlighting the differences caused by including the third dimension.

Moving on to the Q3D simulations results, Figure 12 illustrates the snapshots of the electron number density and electric potential at $t = 30\omega_{pe}$, from Q3D simulations with different numbers of regions and from the full-3D simulation. In addition, Figure 13 provides the time-averaged plasma profiles along the $x$-axis at $y = L_y/2$ and $z = L_z/2$ and along the $z$-axis at $x = 3L_x/4$ and $y = L_y/2$.

The results from Figure 12 and Figure 13 demonstrate that the Q3D simulations utilizing 10 and 25 regions can very closely reproduce the outcomes from the full-3D simulation. Specifically, the snapshots in Figure 12 reveal that these approximations successfully capture the detailed spatial structure of the instability. The time-averaged profiles from the 10- and 25-region simulations are closely aligned with those from the full -3D results. In terms of the temporal behavior, the highly consistent time evolutions of the total electrons' kinetic energy and the electric potential energy resolved by these Q3D simulations compared to 3D, as shown through the final figure in Appendix C.b, reinforces the accurate representation of the underlying physical phenomena.

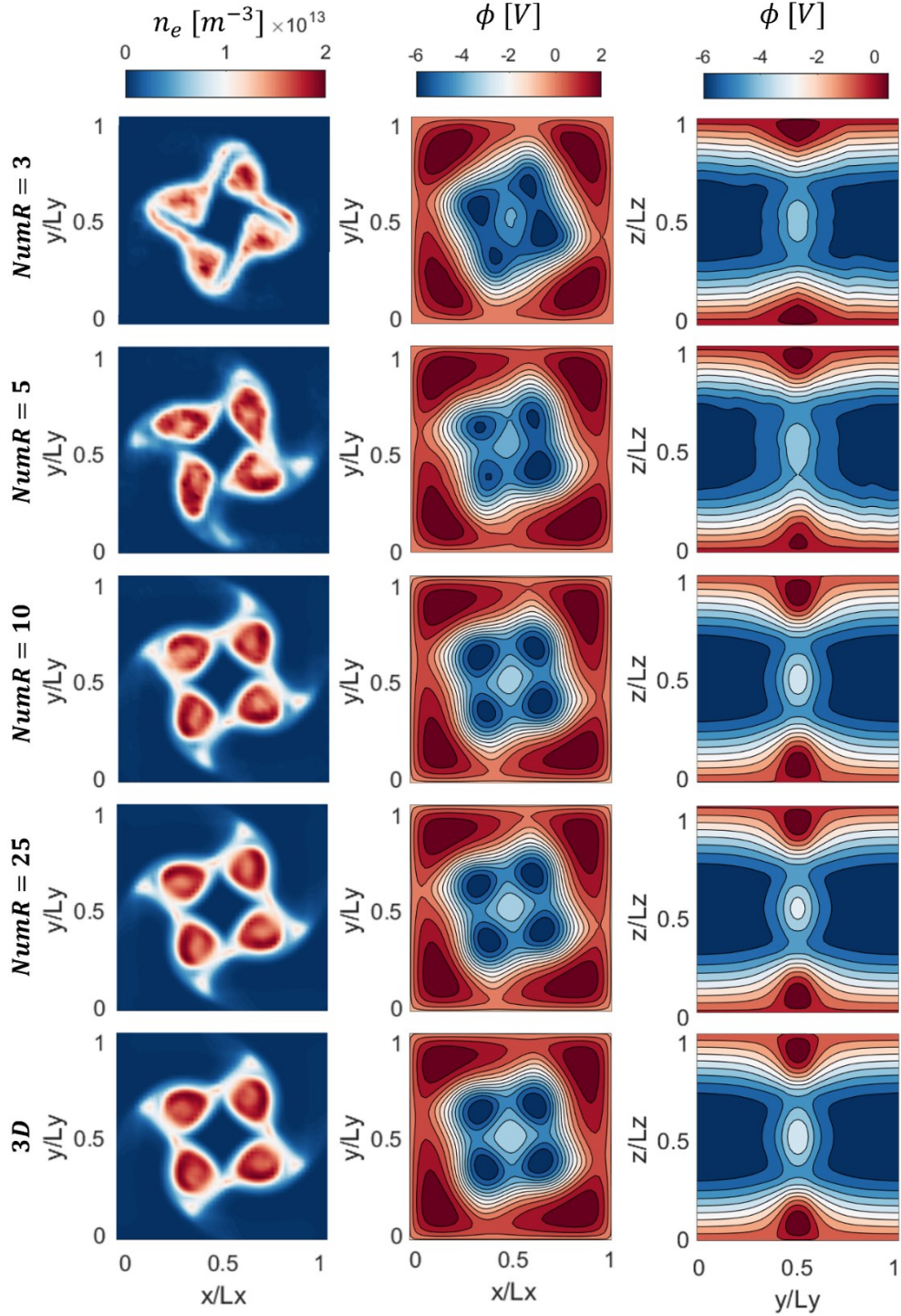

Figure 12: 2D snapshots of plasma density and electric potential from the Q3D and full-3D simulations of test case 2 at time instant of $t = 30\omega_{pe}$; (**left column**) electron number density ($n_e$) in the $y - x$ mid-plane, (**middle column**) plasma potential ($\phi$) in the $y - x$ mid-plane, and (**right column**) plasma potential ($\phi$) the $z - x$ mid-plane.

Even with a reduced number of regions – 3 and 5 – the Q3D simulation retains the essential features and behavior of the system. The approximation provided by the simulation with 5 regions could be acceptable as it only exhibits minor deviations in certain details compared to the full-3D results. However, the simulation with 3 regions shows greater deviations in both the resolved instability structure (Figure 12) and the time-averaged profiles (Figure 13).

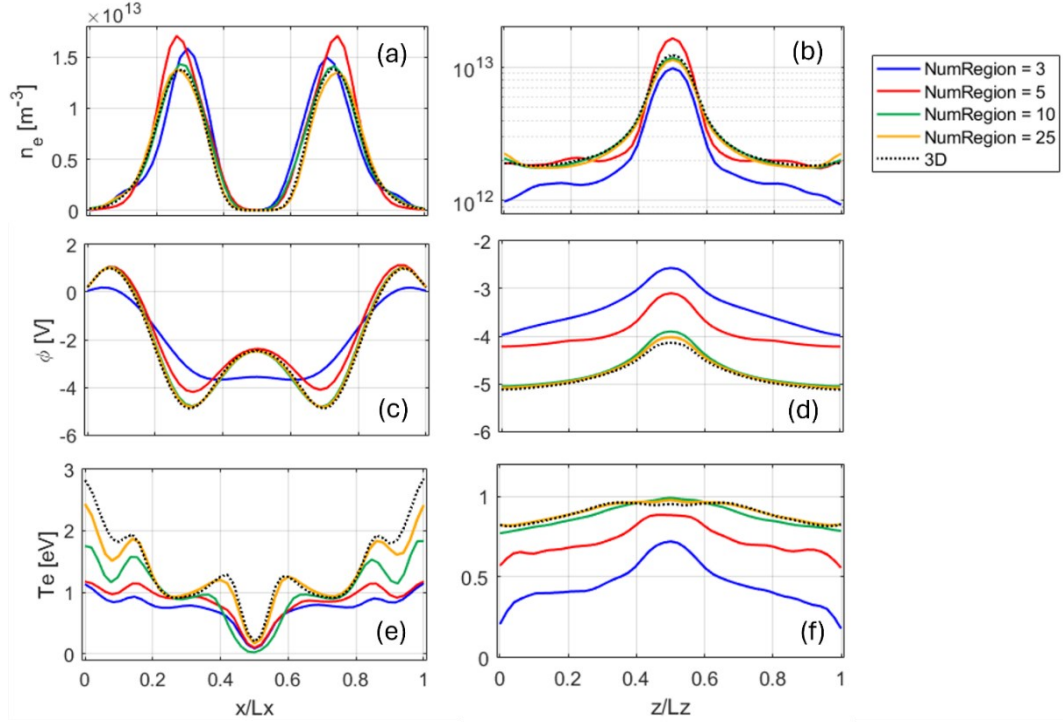

Figure 13: Time-averaged (over $0 - 60\ \omega_{pe}$) profiles of various plasma properties from the Q3D simulations with various number of regions, and from the full-3D simulation of test case 2. The left column represents the profiles along the $x$-direction ($y = L_y/2$ and $z = L_z/2$) and the right column shows the distributions along the z-direction ($x = 3L_x/4$ and $y = L_y/2$); (a) and (b) electron number density ($n_e$), (c) and (d) plasma potential ($\phi$), and (e) and (f) electron temperature ($T_e$).

Looking at Figure 14, we can quantitatively compare the errors in the predictions of the Q3D simulations with various number of regions. The errors are calculated using the L2 norm, measuring the normalized (with respect to the full-3D results) difference between the time-averaged plasma profiles from the Q3D simulations and those from the full-3D.

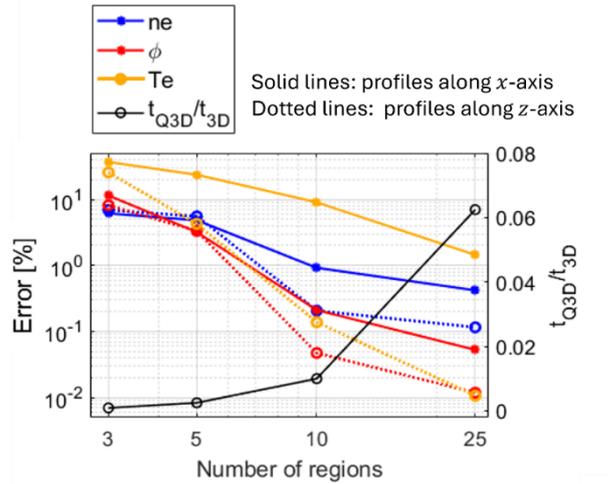

Figure 14: Variation of the L2 error in predictions of the plasma profiles from the Q3D simulations with various number of regions compared to full-3D results in test case 2. The computational time ratio between the Q3D and full-3D simulations is also plotted against the number of regions in the Q3D simulations.

For the Q3D simulation with 25 regions, the errors in various plasma profiles range from 0.01 % to 2 %, while the computational time is only 6 % of that for the full-3D simulation. With 5 regions, the errors in all plasma profiles remain below 5 %, except for the electron temperature profile along the $x$-direction, which exhibits an error of approximately 20 %. This performance is achieved with only 0.25 % computational cost of the full-3D simulation.

It is important to recognize that, any reduced-order model achieves its computational efficiency by introducing certain simplifications, which inevitably lead to some degree of accuracy sacrifice. Therefore, when assessing the errors associated with the RO-PIC, it is important to do so in the context of the computational benefits it offers. The trade-off between computational speedup and accuracy is an inherent aspect of RO-PIC as a reduced-order model, and any evaluation should weigh the accuracy metrics against the computational gain.

The compromise between the accuracy and computational efficiency must be also viewed in light of the specific study's purpose, and the limitations of conventional models under practical computational constraints. When aiming to simulate real plasma systems with standard approaches, one is often forced to make drastic compromises as well. These may include simplifying the problem to 2D [27][28], downscaling the system geometry [29], or greatly limiting the simulation timeframe [3]. While these adjustments can help manage computational demands, they are not always feasible or optimal for capturing the full-picture dynamics of the system. In particular, opting for 2D and losing one dimension altogether often results in the loss of key 3D physics that are essential to the system's behavior.

In contrast, despite creating approximations to the problem, the 3D RO-PIC offers remarkable computational speedups. This allows simulations to maintain the 3D nature of the problem without resorting to downscaling or reducing the problem to 2D. Given the inherently 3D behavior of many plasma systems – including the $E \times B$ plasma technologies – RO-PIC's ability to approximate 3D dynamics, even at lower accuracy, can be far more effective than reducing the problem to 2D. It is highly likely with a 3D RO-PIC simulation, which can be several orders of magnitude faster than conventional PIC methods, we would be able to achieve a closer representation of a 3D plasma system than a highly accurate but 2D model. The rigorous assessment of this assertion is the focus of our upcoming works, where we aim to compare the predictions of a Q3D plasma simulation with experimental data.

In essence, the true advantage of 3D RO-PIC lies in its ability to preserve all three dimensions, and the critical 3D aspects of the system's physics that may otherwise be compromised or distorted when using traditional methods constrained by computational limits.

**Section 5: Conclusions**

In this part II, we extended the first-order RO-PIC scheme introduced in part I [22] to three dimensions, formulating the quasi-3D (Q3D) PIC approach. The Q3D PIC offers a computationally efficient means to simulate three-dimensional plasma dynamics, which are often inaccessible with traditional full-scale 3D PIC approaches due to prohibitive computational costs. The Q3D implementation was verified in this paper through standalone Poisson problems and plasma test cases. The plasma cases comprised the problem of electron plasma oscillations with Landau damping, and the problem of Diocotron instability.

In standalone Poisson test cases, the RDPS showed excellent agreement with full-3D solutions, using two orders of magnitude fewer computational nodes that that needed in the full-3D Poisson solver. Verification results from the two plasma test cases demonstrated that the 3D RO-PIC code captures essential plasma dynamics and key characteristics, such as oscillation frequencies, damping rates, and instability growth, with minimal error. In both test cases, the RO-PIC led to substantial reduction in computations, offering speedup factors of up to two orders of magnitude while preserving involved 3D physics. For example, the characteristics of the electron plasma oscillations, including frequency and damping rate, were captured with error margins below 1 % while achieving a computational cost reduction to just 5 % of that required for a conventional 3D simulation. Similarly, the Diocotron instability case showed that plasma profiles from Q3D simulations presented errors below 2 % at only 6 % of the computational time compared to the corresponding full-3D simulation.

A key strength of the Q3D RO-PIC is its ability to preserve critical three-dimensional aspects of plasma dynamics that are often missed when conventionally reducing simulations to two dimensions or downscaling geometries. The inherently three-dimensional nature of many plasma systems, especially in cross-field configurations, demands accounting for all three dimensions.

The Q3D RO-PIC offers a broad range of trade-offs between accuracy and computational efficiency, making it a versatile tool for various types of plasma simulations. When assessing the accuracy of RO-PIC simulations, it is essential not only to evaluate how closely they match the fully resolved model but also to consider the significant computational savings achieved. In many cases, achieving 90 % of the accuracy of a full-3D simulation with only a fraction of the computational cost represents an exceptionally efficient use of resources. The trade-off between accuracy and efficiency should thus be assessed based on the specific objectives of the simulation.

At lower levels of accuracy, using fewer regions, the Q3D RO-PIC can yield gains in computational efficiency of up to three to four orders of magnitude. This compromise is often suitable for broad exploratory and preliminary studies, where the goal is to gain qualitative or semi-quantitative insights rather than precise, detailed predictions. For more accurate approximations, using higher number of regions, the speedup remains substantial enough to enable rigorous physics studies, as well as precise quantitative predictions.

As the final remark, the 3D RO-PIC method represents a significant advancement in computational tools for plasma physics, bridging the gap between computational feasibility and modeling fidelity. The Q3D PIC opens new possibilities for simulating three-dimensional, large-scale plasma systems that were previously beyond reach due to computational constraints. Future work will focus on further validating the Q3D PIC against experimental data and applying the method to increasingly complex plasma configurations, including those involving turbulence, transport phenomena, and magnetic confinement.

**Appendix**

**A. Zeroth-order Q3D PIC implementation: review of the associated RD Poisson's equations**

This appendix presents the formulation that underpins a zeroth-order Q3D RO-PIC. Before that, we highlight the differences existing between a zeroth-order implementation and the first-order formulation that was discussed in Section 2.

Key distinctions include the weighting functions for the scatter and gather functions, as well as how the 3D Poisson's equation is approximated. The formulations for these are presented below. As a visual aid for the upcoming formulations, the grid system associated with the zeroth-order Q3D formulation, comprising $M \times N \times P$ regions, is illustrated in Figure 15.

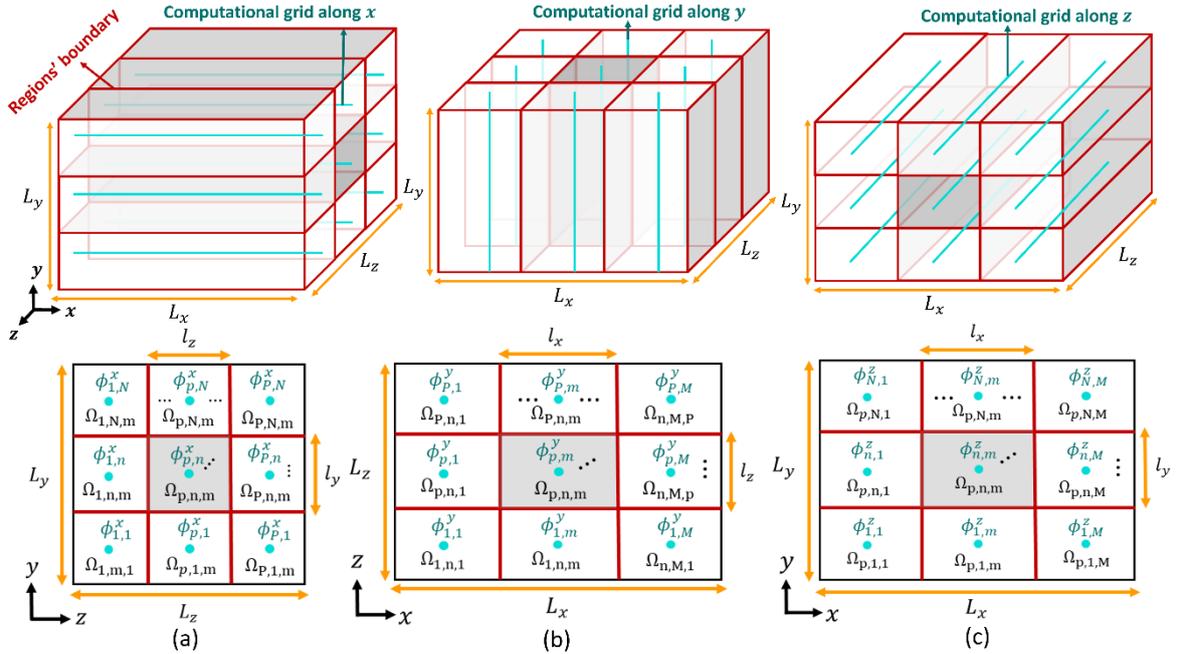

Figure 15: Schematic of the Q3D grid and the domain decomposition into multiple regions for the zeroth-order RO-PIC. The illustrations on the top row demonstrate the 3D views of domain decomposition (including the regions' boundaries and the computational grids) along (a) the $x$-axis, (b) the $y$-axis, and (c) the $z$-axis. The illustrations on the bottom row show the cross-sections of the domain along (a) the $y - z$ plane, (b) the $z - x$ plane (b), and (c) the $x - y$ plane.

The zeroth-order weighting coefficients used to exchange data between the particles and the Q3D grid are

$$W^x_{p,n,i} = \begin{cases} \left(1 - \left|\frac{x - x_i}{\Delta x}\right|\right), & x, y, z \in \Omega_{p,n,m} \\ 0, & x, y, z \notin \Omega_{p,n,m} \end{cases}, \quad \text{(Eq. A1)}$$

$$W^y_{p,m,j} = \begin{cases} \left(1 - \left|\frac{y - y_j}{\Delta y}\right|\right), & x, y, z \in \Omega_{p,n,m} \\ 0, & x, y, z \notin \Omega_{p,n,m} \end{cases}, \quad \text{(Eq. A2)}$$

$$W_{n,m,k}^z = \begin{cases} \left(1 - \left|\frac{z - z_k}{\Delta z}\right|\right), & x, y, z \in \Omega_{p,n,m} \\ 0, & x, y, z \notin \Omega_{p,n,m} \end{cases}.$$ (Eq. A3)

The zeroth-order Q3D Poisson's solution within each region $\Omega_{p,n,m}$, where $m \in [1, M]$, $n \in [1, N]$ and $p \in [1, P]$, is approximated as the direct sum of three potential functions $\phi_{p,n}^x$, $\phi_{p,m}^y$ and $\phi_{n,m}^z$

$$\phi(x, y, z) \approx \phi_{p,n}^x(x) + \phi_{p,m}^y(y) + \phi_{n,m}^z(z), \quad \forall x, y, z \in \Omega_{p,n,m}.$$ (Eq. A4)

Substituting Eq. A4 into Eqs 9-11 yields the following coupled system of equations

$$\left(\frac{d^2\phi_{p,n}^x}{dx^2}\right) l_y l_z + \frac{l_z}{\Delta x} \int_{y_S}^{y_N} \left(\frac{\partial \phi_{p,m}^y}{\partial x}\bigg|_{x_E} - \frac{\partial \phi_{p,m}^y}{\partial x}\bigg|_{x_W}\right) dy + \frac{l_y}{\Delta x} \int_{z_B}^{z_F} \left(\frac{\partial \phi_{n,m}^z}{\partial x}\bigg|_{x_E} - \frac{\partial \phi_{n,m}^z}{\partial x}\bigg|_{x_W}\right) dz$$
$$+ l_z \left[\left(\frac{\partial \phi_{p,n}^x}{\partial y} + \frac{\partial \phi_{p,m}^y}{\partial y} + \frac{\partial \phi_{n,m}^z}{\partial y}\right)_{y_N} - \left(\frac{\partial \phi_{p,n}^x}{\partial y} + \frac{\partial \phi_{p,m}^y}{\partial y} + \frac{\partial \phi_{n,m}^z}{\partial y}\right)_{y_S}\right]$$
$$+ l_y \left[\left(\frac{\partial \phi_{p,n}^x}{\partial z} + \frac{\partial \phi_{p,m}^y}{\partial z} + \frac{\partial \phi_{n,m}^z}{\partial z}\right)_{z_F} - \left(\frac{\partial \phi_{p,n}^x}{\partial z} + \frac{\partial \phi_{p,m}^y}{\partial z} + \frac{\partial \phi_{n,m}^z}{\partial z}\right)_{z_B}\right]$$
$$= -\frac{1}{\epsilon_0} \int_{z_B}^{z_F} \int_{y_S}^{y_N} \rho(x, y, z) dy dz,$$ (Eq. A5)

$$\left(\frac{d^2\phi_{p,m}^y}{dy^2}\right) l_x l_z + \frac{l_z}{\Delta y} \int_{x_W}^{x_E} \left(\frac{\partial \phi_{p,n}^x}{\partial y}\bigg|_{y_N} - \frac{\partial \phi_{p,n}^x}{\partial y}\bigg|_{y_S}\right) dx + \frac{l_x}{\Delta y} \int_{z_B}^{z_F} \left(\frac{\partial \phi_{n,m}^z}{\partial y}\bigg|_{y_N} - \frac{\partial \phi_{n,m}^z}{\partial y}\bigg|_{y_S}\right) dz$$
$$+ L_{z,p} \left[\left(\frac{\partial \phi_{p,n}^x}{\partial x} + \frac{\partial \phi_{p,m}^y}{\partial x} + \frac{\partial \phi_{n,m}^z}{\partial x}\right)_{x_E} - \left(\frac{\partial \phi_{p,n}^x}{\partial x} + \frac{\partial \phi_{p,m}^y}{\partial x} + \frac{\partial \phi_{n,m}^z}{\partial x}\right)_{x_W}\right]$$
$$+ l_x \left[\left(\frac{\partial \phi_{p,n}^x}{\partial z} + \frac{\partial \phi_{p,m}^y}{\partial z} + \frac{\partial \phi_{n,m}^z}{\partial z}\right)_{z_F} - \left(\frac{\partial \phi_{p,n}^x}{\partial z} + \frac{\partial \phi_{p,m}^y}{\partial z} + \frac{\partial \phi_{n,m}^z}{\partial z}\right)_{z_B}\right]$$
$$= -\frac{1}{\epsilon_0} \int_{z_B}^{z_F} \int_{x_W}^{x_E} \rho(x, y, z) dx dz,$$ (Eq. A6)

$$\left(\frac{d^2\phi_{n,m}^z}{dz^2}\right) l_x l_y + \frac{l_x}{\Delta z} \int_{y_S}^{y_N} \left(\frac{\partial \phi_{p,m}^y}{\partial z}\bigg|_{z_F} - \frac{\partial \phi_{p,m}^y}{\partial z}\bigg|_{z_B}\right) dy + \frac{l_y}{\Delta z} \int_{x_W}^{x_E} \left(\frac{\partial \phi_{p,n}^x}{\partial z}\bigg|_{z_F} - \frac{\partial \phi_{p,n}^x}{\partial z}\bigg|_{z_B}\right) dx$$
$$+ l_y \left[\left(\frac{\partial \phi_{p,n}^x}{\partial x} + \frac{\partial \phi_{p,m}^y}{\partial x} + \frac{\partial \phi_{n,m}^z}{\partial x}\right)_{x_E} - \left(\frac{\partial \phi_{p,n}^x}{\partial x} + \frac{\partial \phi_{p,m}^y}{\partial x} + \frac{\partial \phi_{n,m}^z}{\partial x}\right)_{x_W}\right]$$
$$+ l_x \left[\left(\frac{\partial \phi_{p,n}^x}{\partial y} + \frac{\partial \phi_{p,m}^y}{\partial y} + \frac{\partial \phi_{n,m}^z}{\partial y}\right)_{y_N} - \left(\frac{\partial \phi_{p,n}^x}{\partial y} + \frac{\partial \phi_{p,m}^y}{\partial y} + \frac{\partial \phi_{n,m}^z}{\partial y}\right)_{y_S}\right]$$
$$= -\frac{1}{\epsilon_0} \int_{y_S}^{y_N} \int_{x_W}^{x_E} \rho(x, y, z) dx dy,$$ (Eq. A7)

**B. Additional Poisson problems for first-order 3D RDPS verification**

The 3D RDPS' solutions for three additional test problems are provided in Figure 16. These test cases differ from each other based on the types of conditions implemented along the boundary surfaces. The source term for these test cases represents a charge distribution given by

$$\rho = z\sin(5\pi y) + exp(-50((y - 0.5)^2 + (z - 0.5)^2))$$ (Eq. A8)

The boundary conditions include: (a) zero-Volt Dirichlet conditions along the $x$ and $y$ directions with a periodic condition along the $z$ direction, (b) zero-Volt Dirichlet conditions along the $x$ and $y$ directions with a Neumann condition along the $z$ direction, and (c) a zero-Volt Dirichlet condition along the $x$ direction, a Neumann condition along the $y$ direction, and a periodic condition along the $z$ direction.

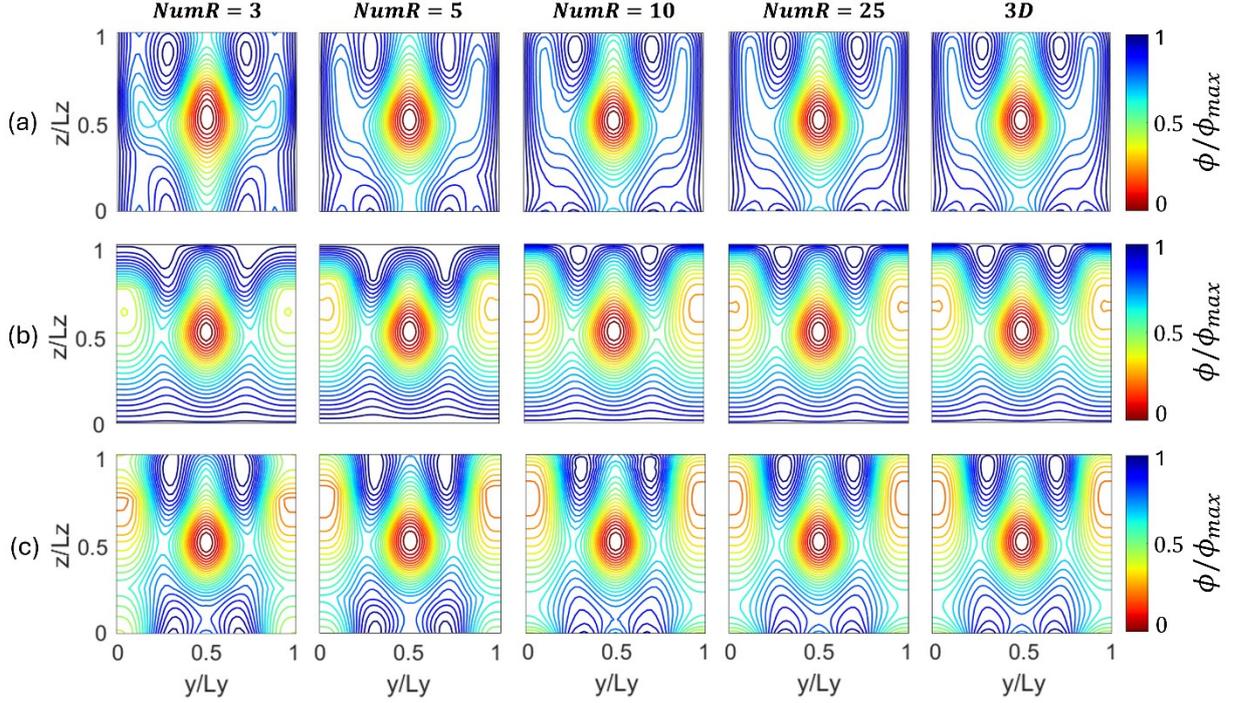

Figure 16: Comparison of the first-order 3D RDPS' solutions with various number of regions against the corresponding full-3D solution (**right-most column**) for additional test cases with (a) Dirichlet boundary conditions along $x$ and $y$ directions and periodic boundary condition along $z$ (b) Dirichlet boundary conditions along $x$ and $y$ directions and Neumann boundary condition along $z$ and (c) Dirichlet boundary conditions along $x$, Neumann boundary condition along $y$ and periodic boundary condition along $z$. The numbers of regions are equal along all three dimensions ($M = N = P$), which are denoted by $NumR$.

## C. Additional results from the plasma test cases

This appendix provides supplementary figures and results from the plasma test cases discussed in the main text, offering further support for the arguments presented.

### a. Test case 1: Electron plasma oscillations

Figure 17 presents a comparison between full-2D and full-3D simulations of test case 1, focusing on the temporal evolution of the system's electric potential energy and the particles' kinetic energy.

The figure depicts the impact of including the third dimension in the problem, in particular with regards to the oscillations amplitude and their damping rate. The 2D simulation exhibits stronger oscillations, while in the 3D configuration, the oscillations' amplitude is reduced, and it diminishes at a faster rate.

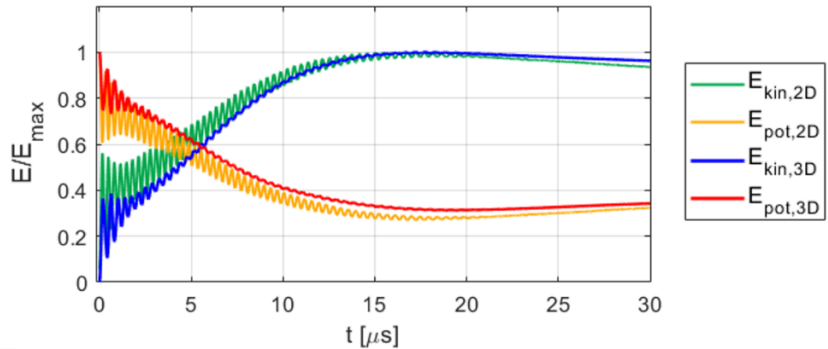

Figure 17: Time evolution of the normalized total electric potential energy ($E_{pot}$) and the total electrons' kinetic energy ($E_{kin}$), from the full-2D and the full-3D simulations of test case 1. The normalization factor ($E_{max}$) is the maximum of the respective energies in the simulation over the simulated duration.

Figure 18 illustrates the snapshots of the electron plasma density from the Q3D simulations with various number of regions, demonstrating how increasing the number of regions affects the simulation resolution. Notably, despite lower resolutions, the core behavior of the system is properly captured even with a few regions.

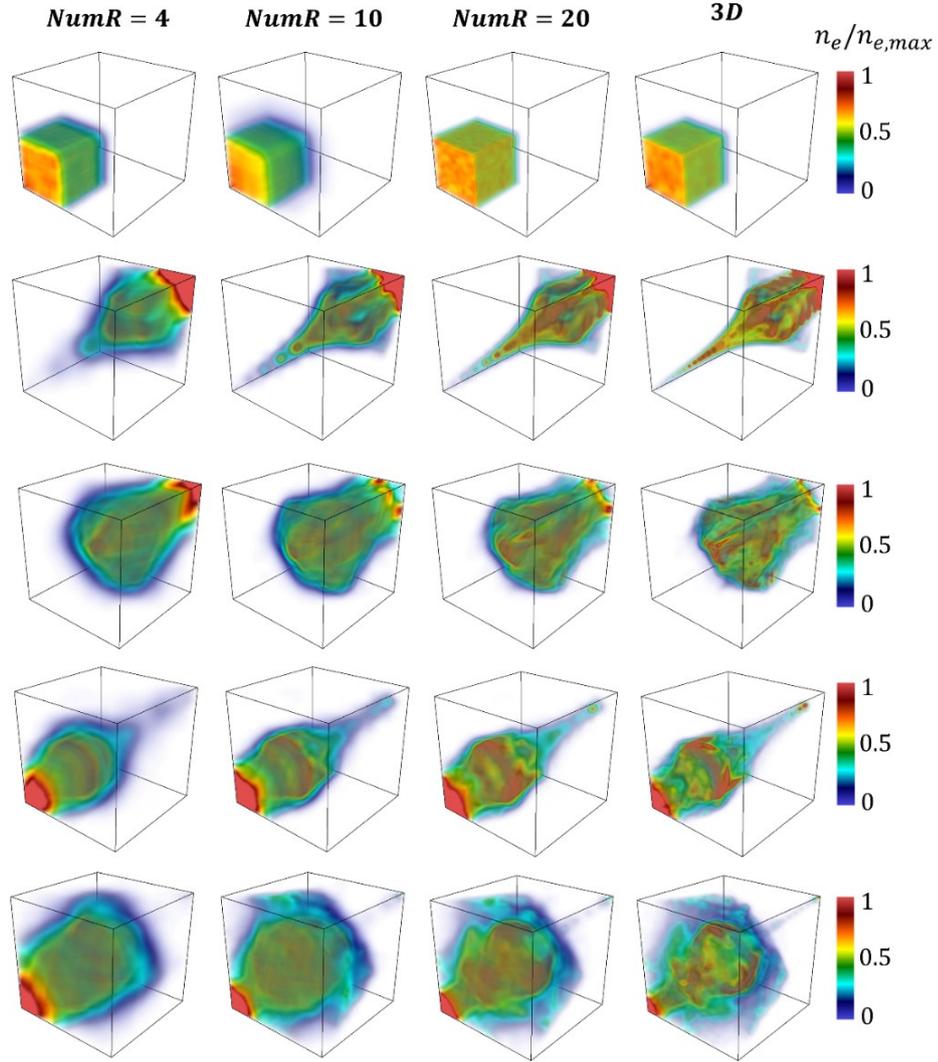

Figure 18: 3D snapshots of the normalized electron number density ($n_e/n_{e,max}$, where $n_{e,max}$ is the peak $n_e$ in the respective snapshot from the full-3D simulation), from the Q3D and the full-3D simulations at five different instants of the plasma oscillations in test case 1.

### b. Test case 2: Diocotron instability

Figure 19 provides 3D visualizations of the electron number density distribution at four distinct time instants during the instability evolution from a full-3D simulation of test case 2. The figure showcases the 3D structure and evolution of the plasma in this case. The snapshots illustrate the development and growth of the instability simultaneously to the dispersion of the electrons along and across the $z$-axis.

A comparison between the time-averaged plasma properties from the full-2D and full-3D simulations, conducted under similar initial conditions, is shown in Figure 20. These plots highlight the impact of the third dimension in the overall plasma distributions.

Finally, Figure 21 presents the time evolution of the total kinetic energy of the electrons and the electric potential energy, as obtained from the Q3D simulations with 10 and 25 regions and compared against those from the full-3D simulation. The figure underscores the accurate capturing of the plasma dynamics with these Q3D approximations.

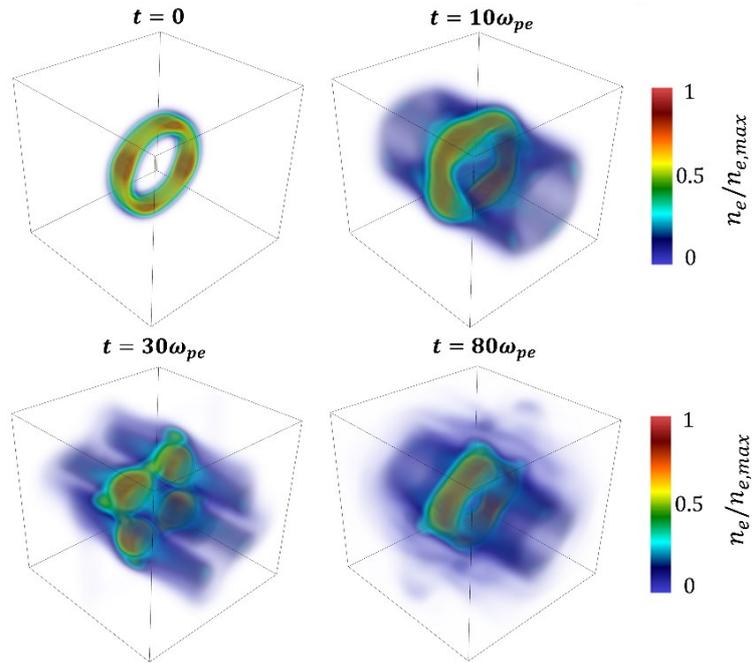

Figure 19: 3D snapshots of the normalized electron number density ($n_e/n_{e,max}$) from the full-3D simulation of test case 2 at four different time instants during the Diocotron instability evolution.

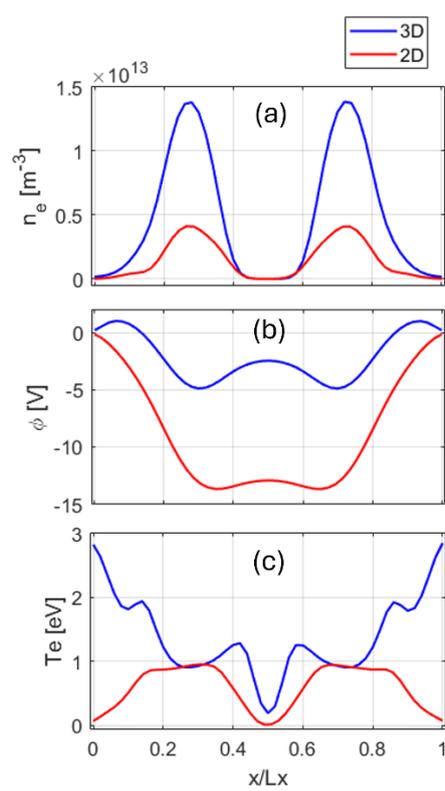

Figure 20: Time-averaged (over $0 - 60\ \omega_{pe}$) profiles of various plasma properties from the full-3D and the full-2D simulations of the Diocotron instability test case. The profiles are plotted along the $x$-direction ($y = L_y/2$ and $z = L_z/2$); (a) electron number density ($n_e$), (b) plasma potential ($\phi$), and (c) electron temperature ($T_e$).

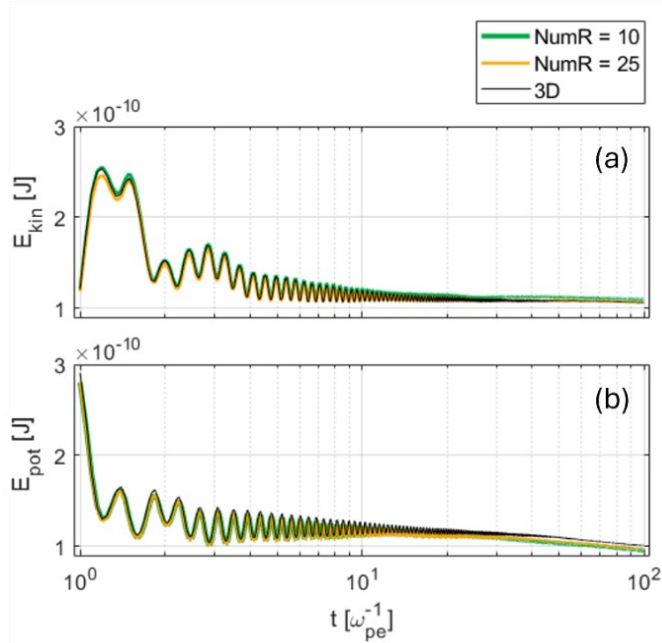

Figure 21: Time evolution of (a) total electrons' kinetic energy ($E_{kin}$) and (b) electric potential energy ($E_{pot}$), from the Q3D (with 10 and 25 regions) simulations and the full-3D simulation of test case 2. The $x$-axes in both plots are in logarithmic scale.